\documentclass[a4paper]{article}

\usepackage[margin=25mm]{geometry}
\usepackage{amsmath}
\usepackage{amsfonts}
\usepackage{amssymb}
\usepackage{graphicx}
\usepackage{verbatim}
\immediate\write18{texcount -tex -sum  \jobname.tex > \jobname.wordcount.tex}

\usepackage{multirow}

\usepackage{subfiles} 

\usepackage{cite}
\usepackage{algorithmic}
\usepackage{textcomp}

\usepackage[dvipsnames]{xcolor}
\usepackage{tikz,graphics,float,epsf,caption,subcaption}
\usepackage{3dplot} 
\usepackage{esvect}
\usepackage{array}
\usepackage{pgfplots}

\usetikzlibrary{calc, intersections, angles}

\providecommand{\keywords}[1]
{
  \small	
  \textbf{\textit{Keywords---}} #1
}

\title{2+2D Texture for Full Positive Parallax Effect}
\author{Dias A. Yip Gonçalves$^{1}$, Zuffo M. Knörich$^{2}$  \\
        \small $^{1}$Dept. of Electronic Systems Engineering, Polytechnic School of the USP, São Paulo, SP - Brazil\\
        \small $^{2}$Dept. of Electronic Systems Engineering, Polytechnic School of the USP, São Paulo, SP - Brazil
}

\begin{document}

\maketitle

\footnote{This study was financed in part by the Coordenação de
Aperfeiçoamento de Pessoal de Nível Superior - Brasil (CAPES) - Grant Number 88887.607921/2021-00}

\begin{abstract}
The representation of parallax on virtual environment is still a problem to be studied. Common algorithms, such as Bump Mapping, Parallax Mapping and Displacement Mapping, treats this problem for small disparity between a ``real'' object and a simplified model. This work will introduce a new texture structure and one possible render algorithm able to display parallax for large disparities, it is an approach based on the four-dimensional representation of the Light Field and was thought to positive parallax and to display the surfaces on the inside of our simplified model. These conditions are imposed to allow the free movement of an observer, if its movement is restrict, these conditions may be loosen. It is a high storage low process approach possible to be used in real time systems. As an example we will develop a scene with several objects and simplified them by a unique sphere that encloses them all, our system was able to run this scene with about 180fps.
\end{abstract}


\section{Introduction}
\label{sec:introduction}
In a virtual environment, textures are images used as a lookup table to apply colour effects in objects. It is a powerful tool able to combine several layers of effects with the finality to bestow realism to a created world, and improve the users' perception of immersion. By being static elements, the way that several textures are combined to generate rendered views is of utmost importance, classical techniques involve the combination of reflection and parallax techniques. The most common reflection techniques treats the diffuse and specular reflection of objects, while parallax is usually associated to bump texture. Bump texture records, instead of colours, the height displacement of a ``real'' object when compared to its simplificated model and is used to produce the illusion of unevenness over a surface, on what is called Bump Mapping\cite{Blinn1978SimulationOW}. \cite{Kaneko2001} was able to improve this technique to add parallax representation and called it Parallax Mapping.

Displacement Mapping\cite{Cook84} is another technique able to display parallax through a bump texture. It consists of changing the simplified model on render time to approximate it to the ``real'' object. All this technique works well for small unevenness, but fail to present satisfactory results when they become more expressive. \cite{wang2003} does a brief analysis of the strong and weak points of each technique considering shadow, occlusion, silhouette and inter-reflection representation, and introduces a five-dimensional approach to the parallax problem. 

This work will introduce a new structure of texture able to show parallax for larger unevenness and even represent transparency for discontinuous objects. This approach was based on the Light Field concept and may be thought as an expansion of the traditional concept of 2D texture, working for any object, and is able to reproduce any interior surface and record any occlusion, shadowing, inter-reflection or transparency noted during its synthesis, reproducing it in real time.

\par The Light Field was first proposed as a seven-dimensional scalar field\cite{adelson1991},
\begin{equation}
\label{eq:plenoptica}
P=P(\theta,\phi,\lambda,t,V_x,V_y,V_z),
\end{equation}
where $V_x$, $V_y$, and $V_z$ determines an observer position, $\theta$ and $\phi$ a direction of light in spheric coordinates, $t$ an instant in time, and $\lambda$ a wavelength. 
\cite{levoy96} simplified this plenoptic function to a four-dimensional expression. They first eliminated the $t$ and $\lambda$ components by working with photographies (fixing an instant and selecting only the red, green, and blue wavelength) and then noted that by pre-determining two planes, a ray may be uniquely determined by the intersection point on these two surfaces, they called the coordinates of the first point spatial dimensions and the second angular dimensions. \cite{gortler96} made similar assumptions and called this four-dimensional representation ``lumigraph''.

Light Field based techniques for treating the parallax problem usually involves image based rendering (IBR), when several photographies from different points of view are used to render a different one through the interpolation of the input pictures or using some artificial intelligence algorithm to fill the missing pixels. This technique is used to produce panoramic views of a scene and is able to change the observer's position until certain limitation\cite{Overbeck2018}\cite{Lyu2020}, but is strongly dependent of the layout of cameras used and lacks a standardized way of being able to reproduce this scenes in any system. 


\par The texture here proposed is a simple texture whose goal is to represent positive parallax of any real surface inside a model object. This structure may be applied to any object inside a virtual world with the only condition that the normal vector must be specified for every point. The positive parallax limitation was imposed to make the free placement of an observer outside our model possible, if some placement restriction were imposed, this system may be used to also represent negative parallax.

\par It is also possible to invert this idea and place an observer inside it, in this case the result will be similar to a panoramic view of a scene.

\par As an example, we will create two virtual scenes with objects and approximate them by only one spheric model object that englobes them all. 

\par
\begin{figure}
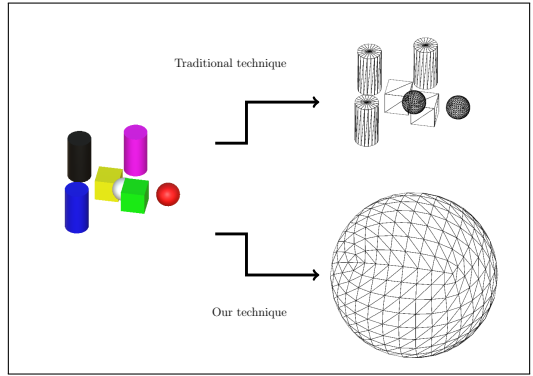

	\fbox{
		\resizebox{0.4\textwidth}{!}{
			\subfile{dias1a}
		}
	}
	\centering
	\caption{Comparative between the traditional techniques' model  and our model.}
	\label{fig:comparative}
\end{figure}

 Note that the discrepancy between ``real'' and model here is far superior than any other technique before mentioned, as showed in fig. \ref{fig:comparative}. The final result produced by our texture will simulate the effect of a ``snow globe'' in the sense that you are able to see the objects in our model's interior through its spheric surface. And may be used as a structure interface for improve a spheric display, such as the OrbeVR\cite{Zuffo2017}, to a true multi-viewer experience without the assistance of any tracking equipment.

\section{2+2D Texture Theory}

Textures are discrete elements of a defined size and are usually represented by the normalization of its dimension in the range 0 to 1. This will also be the representation used in this work. 
\par As mentioned before our texture structure will be based in the four-dimensional representation of the light field and expands the usual concept of two dimensional texture by two dimensions, the angular dimensions. We can represent our texture by
\begin{equation}
\label{eq:tex_exp}
T=T(u,v,s,t),
\end{equation}
where $T$ stores the value of our pixel texture, $u$ and $v$ represents the spatial dimension and works like the usual uv mapping of a two dimensional texture over a surface ((\ref{eq:sphere_par}) shows a parametrization for a spheric surface), and $s$ and $t$ are the angular dimension, producing a texture with a total of four dimensions of size $U$, $V$, $S$ and $T$, that represents the equivalent light ray of our scene as emitted by our model object' surface.

\begin{equation}
	\begin{cases}
		u = \frac{\theta+180^{\circ}}{360^{\circ}} = \frac{atan2(P_x,P_z)+180^{\circ}}{360^{\circ}}\\
		v = \frac{\phi+90^{\circ}}{180^{\circ}} = \frac{asin(P_y/\lvert\lvert\vv{OP}\rvert\rvert)+90^{\circ}}{180^{\circ}}
	\end{cases}
	\label{eq:sphere_par}
\end{equation}

\par
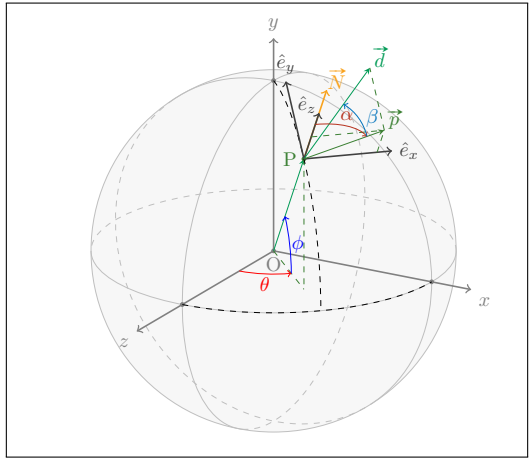
\begin{figure}
	\fbox{
		\resizebox{.4\textwidth}{!}{
				\centering
	
	\pgfmathsetmacro{\thetad}{70}
	\pgfmathsetmacro{\phid}{120}

	\pgfmathsetmacro{\rvec}{.8}
	\pgfmathsetmacro{\thetavec}{40}
	\pgfmathsetmacro{\phivec}{45}

	\pgfmathsetmacro{\rvecb}{.5}
	\pgfmathsetmacro{\thetavecb}{50}
	\pgfmathsetmacro{\phivecb}{60}

	\tdplotsetmaincoords{\thetad}{\phid}
		
	\begin{tikzpicture}[scale=4,tdplot_main_coords]
		\definecolor{superlightgray}{rgb}{.97, .97, .97}
		\filldraw[fill = superlightgray, draw = lightgray, style={x={(1 cm,0 cm)},y={(0 cm, 1 cm)},z={(0 cm, 0 cm)}}] (0,0) circle (\rvec); 

		\coordinate (O) at (0,0,0);
		
		\tdplotsetrotatedcoords{\phid-180}{0}{0}%
		\draw[tdplot_rotated_coords, ultra thin, color= lightgray] (\rvec,0,0) arc (0:180:\rvec);
		\tdplotresetrotatedcoordsorigin
		\tdplotsetrotatedcoords{\phid}{0}{0}%
		\draw[tdplot_rotated_coords,ultra thin, dashed, color= lightgray] (\rvec,0,0) arc (0:180:\rvec);
		\tdplotsetrotatedcoords{0}{0}{0}%
		\tdplotSetRotatedAlphaPlaneCoords{90}	
		\tdplotSetRotatedBetaPlaneCoords{-40}
		\draw[tdplot_rotated_coords, ultra thin, color= lightgray] (\rvec,0,0) arc (0:180:\rvec);
		\tdplotSetRotatedBetaPlaneCoords{180}
		\draw[tdplot_rotated_coords,ultra thin, dashed, color= lightgray] (\rvec,0,0) arc (0:180:\rvec);
		\tdplotsetrotatedcoords{0}{0}{0}%
		\tdplotSetRotatedBetaPlaneCoords{-20}
		\draw[tdplot_rotated_coords, ultra thin, color= lightgray] (\rvec,0,0) arc (0:180:\rvec);
		\tdplotSetRotatedBetaPlaneCoords{180}
		\draw[tdplot_rotated_coords,ultra thin, dashed, color= lightgray] (\rvec,0,0) arc (0:180:\rvec);

		\tdplotsetcoord{P}{\rvec}{\thetavec}{\phivec}


		\draw[thick,->, gray] (0,0,0) -- (1.2,0,0) node[anchor=north east]{$z$};
		\draw[thick,->, gray] (0,0,0) -- (0,1,0) node[anchor=north west]{$x$};
		\draw[thick,->, gray] (0,0,0) -- (0,0,1) node[anchor=south]{$y$};
		
		\fill[color=gray] (O) circle[radius=0.3pt] node[anchor = north] {O};
		
		\fill[gray] (\rvec,0,0) circle (.3pt);
		\fill[gray] (0,\rvec,0) circle (.3pt);
		\fill[gray] (0,0,\rvec) circle (.3pt);
		
		\draw[-stealth,color=ForestGreen] (O) -- (P);

		\draw[dashed, color=OliveGreen] (O) -- (Pxy);
		\draw[dashed, color=OliveGreen] (P) -- (Pxy);

		\fill[color=OliveGreen] (P) circle[radius=0.3pt] node[anchor = east] {P};

		\tdplotdrawarc[->, red]{(O)}{0.3}{0}{\phivec}{yshift=-5pt}{$\theta$}

		\tdplotsetthetaplanecoords{\phivec}

		\tdplotdrawarc[tdplot_rotated_coords, <-, blue]{(0,0,0)}{.3}{\thetavec}{90}{xshift=4pt}{$\phi$}

		\draw[dashed,tdplot_rotated_coords] (\rvec,0,0) arc (0:90:\rvec);
		\draw[dashed] (\rvec,0,0) arc (0:90:\rvec);

		\tdplotsetrotatedcoords{\phivec}{\thetavec-90}{0}
		

		\tdplotsetrotatedcoordsorigin{(P)}

		\draw[thick,tdplot_rotated_coords,->, YellowOrange] (0,0,0) -- (.6,0,0) node[xshift=5pt, yshift=5pt]{$\vv{N}$};
		\draw[thick,tdplot_rotated_coords,->, darkgray] (0,0,0) -- (.4,0,0) node[xshift=-6pt, yshift=4pt]{$\hat{e}_z$};
		\draw[thick,tdplot_rotated_coords,->, darkgray] (0,0,0) -- (0,.4,0) node[anchor=west]{$\hat{e}_x$};
		\draw[thick,tdplot_rotated_coords,->, darkgray] (0,0,0) -- (0,0,.4) node[anchor=south]{$\hat{e}_y$};


		\pgfmathsetmacro{\xnew}{\rvecb*cos(\phivecb)*sin(\thetavecb)}
		\pgfmathsetmacro{\ynew}{\rvecb*sin(\phivecb)*sin(\thetavecb)}
		\pgfmathsetmacro{\znew}{\rvecb*cos(\thetavecb)}

		\draw[-stealth,color=ForestGreen,tdplot_rotated_coords] (0,0,0) -- (\xnew,\ynew,\znew) node[xshift=5pt, yshift=5pt]{$\vv{d}$};
		\draw[-stealth,color=OliveGreen,tdplot_rotated_coords] (0,0,0) -- (\xnew,\ynew,0) node[xshift=5pt, yshift=5pt]{$\vv{p}$};
		\draw[dashed,color=OliveGreen,tdplot_rotated_coords] (\xnew,\ynew,0) -- (\xnew,\ynew,\znew);
		\draw[dashed,color=OliveGreen,tdplot_rotated_coords] (\xnew,0,0) -- (\xnew,\ynew,0);
		\draw[dashed,color=OliveGreen,tdplot_rotated_coords] (0,\ynew,0) -- (\xnew,\ynew,0);
		
		\tdplotdrawarc[tdplot_rotated_coords,color=BrickRed, ->]{(0,0,0)}{0.3}{0}{\phivecb}{yshift=5pt}{$\alpha$}

		\tdplotsetrotatedcoords{\phivec}{\thetavec}{90-\phivecb}
		
		\tdplotdrawarc[tdplot_rotated_coords,color=RoyalBlue, ->]{(0,0,0)}{0.3}{90}{180-\thetavecb}{xshift=7pt}{$\beta$}
	\end{tikzpicture}
	
		}
	}
	\centering
	\caption{Coordinate system used to parametrize the four-dimensional texture.}
	\label{fig:parametrization}
\end{figure}

 Fig. \ref{fig:parametrization} illustrates the parametrization used here. $P$ is a generic point on our surface. Through its normal vector ($\vv{N}$) and the global up vector we are able to define a unique orthogonal local coordinate system for each point in our surface with its axis defined by its normal vector ($\hat{e}_z$), a vector pointing to the global up vector ($\hat{e}_y$) and a vector orthogonal to both ($\hat{e}_x$). This new coordinate system is specially interesting because we are able to use its plane x-y to easily divide the space in two hemisphere, an internal to the object and an external, and as we intend to position the observer strictly outside our surface, the $s$ and $t$ dimensions maps only to range $-90^{\circ}$ to $90^{\circ}$.

Using this local system, we are able to uniquely parametrize the light direction ($\vv{d}$) of a ray that passes through $P$ by its spherical coordinates in our new coordinate system. In summary, the local coordinate system may be determined by:
\begin{equation}
	\begin{cases}
		\hat{e}_z = \hat{N} \\
		\hat{e}_x = \frac{\hat{y} \times \hat{e}_z}{\lvert\lvert \hat{y} \times \hat{e}_z\rvert\rvert} \\
		\hat{e}_y = \hat{e}_z \times \hat{e}_x
	\end{cases}.
	\label{eq:second_coord_system}
\end{equation}
And $s$ and $t$ are calculated by:
\begin{equation}
	\begin{cases}
		s = \frac{\alpha+90^{\circ}}{180^{\circ}} = \frac{asin(\hat{e}_x^T\hat{p})+90^{\circ}}{180^{\circ}}\\
		t = \frac{\beta+90^{\circ}}{180^{\circ}} = \frac{asin(\hat{e}_y^T\hat{d})+90^{\circ}}{180^{\circ}}
	\end{cases},
	\label{eq:angula_dimension}
\end{equation}
where $\hat{p}$ is the versor projection of $\hat{d}$ over the plane formed by the versors $\hat{e}_z$ and $\hat{e}_x$.

\par By uniting \ref{eq:sphere_par} and \ref{eq:angula_dimension} we are able to uniquely describe any light ray that pass through and leaves our spheric surface in the range 0 to 1 as before mentioned.

\section{Render Process}

Considering that a texture is a discrete element of size relatable to its resolution, not all points over a continuous surface will match perfectly this discrete object. For these points it’s important to define an algorithm that mix some of the texture’s value and finds an approximate colour value to screen pixels for them. For 2D textures the renderization usually involves the use of bilinear filter to interpolate the closest pixel values. This same idea will be applied to our spatial dimensions. 
\par However the closest pixel values may only be determined by taking in consideration the angular dimensions. For the angular dimensions we propose also the use of bilinear interpolation, but this time over the angles that takes into consideration the same direction view of the desired point to the observer in their local coordinate system, for all points used for the interpolation of the spatial dimensions.
\par
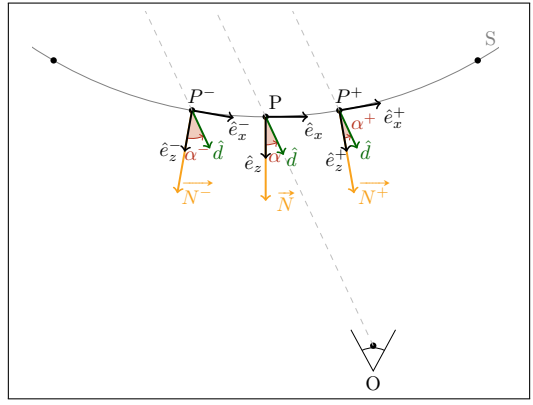
\begin{figure}
	\fbox{
		\resizebox{0.4\textwidth}{!}{



\begin{tikzpicture}[scale=.75]
	\usetikzlibrary{calc}
	\usetikzlibrary {angles,quotes}
	
	\clip (-5.5,-7.5) rectangle + (11,-9);

	\pgfmathsetmacro {\sizeTex}{18}
	\pgfmathsetmacro {\point} {13.5}
	\pgfmathsetmacro {\obsAngle} {-65}
	
	\pgfmathparse{cos(\obsAngle)}
	\pgfmathsetmacro {\value}{\pgfmathresult}
	\pgfmathparse{sin(\obsAngle)}
	\coordinate (dir) at (\value, \pgfmathresult);

	\draw[line width=0.15, color=gray] (0,0) circle (10);
	\foreach \x in {1,...,\sizeTex}
	{
		\pgfmathparse{10*cos(360/\sizeTex*\x)}
		\pgfmathsetmacro {\PvalueX}{\pgfmathresult}
		\pgfmathparse{10*sin(360/\sizeTex*\x)}
		\fill (\PvalueX,\pgfmathresult) circle(0.075);
	}	
	
	\pgfmathparse{10*cos(360/\sizeTex*15.1)}
	\pgfmathsetmacro {\PvalueX}{\pgfmathresult}
	\pgfmathparse{10*sin(360/\sizeTex*15.1)}
	\coordinate[label={[xshift=.1, yshift=0.3, color=gray]S}] (S) at (\PvalueX,\pgfmathresult);
	
	\pgfmathparse{10*cos(360/\sizeTex*\point)}
	\pgfmathsetmacro {\Px}{\pgfmathresult}
	\pgfmathparse{10*sin(360/\sizeTex*\point)}
	\pgfmathsetmacro {\Py}{\pgfmathresult}
	\fill (\Px,\Py) circle (0.075) coordinate[label={[xshift=5, yshift=0.3]P}](P);

	\fill ($(P)+6*(dir)$) circle (.075) coordinate(O);
	\draw[dashed, color=lightgray] (O) -- (P);
	\draw ($(O)+(0,-0.05)$) arc (90:64:0.6) coordinate(Or);
	\draw ($(O)+(0,-0.05)$) arc (90:116:0.6) coordinate(Ol);
	\coordinate[label={[xshift=0, yshift=-13]O}] (Ob) at ($(O)+(0,-0.6)$);
	\path[draw] ($2*(Ol)-(Ob)$) -- (Ob) -- ($2*(Or)-(Ob)$);

	\pgfmathsetmacro {\pointminus}{floor(\point)}
	\pgfmathparse{\pointminus+1}
	\pgfmathsetmacro {\pointplus}{\pgfmathresult}
	
	{
		\pgfmathparse{cos(360/\sizeTex*\pointminus)}
		\pgfmathsetmacro {\PvalueX}{\pgfmathresult}
		\pgfmathparse{sin(360/\sizeTex*\pointminus)}
		\pgfmathsetmacro {\PvalueY}{\pgfmathresult}
		\coordinate[label={[xshift=5, yshift=0.3]$P^-$}] (Pw) at ($10*(\PvalueX,\PvalueY)$);
		\coordinate[label={[xshift=-7, yshift=-8]$\hat{e}^-_z$}] (No) at ($11*(\PvalueX,\PvalueY)$);
		
		\pgfmathparse{cos(360/\sizeTex*\pointminus+90)}
		\pgfmathsetmacro {\TvalueX}{\pgfmathresult}
		\pgfmathparse{sin(360/\sizeTex*\pointminus+90)}
		\pgfmathsetmacro {\TvalueY}{\pgfmathresult}
		\coordinate[label={[xshift=3, yshift=-15]$\hat{e}^-_x$}] (To) at ($(Pw)+(\TvalueX,\TvalueY)$);
		
		\coordinate[label={[xshift=4, yshift=-9, color=black!60!green]$\hat{d}$}] (Do) at ($(Pw)+(dir)$);
		
		\draw[->, line width=1, color=YellowOrange] (Pw)->($2*(No)-(Pw)$) coordinate[label={[xshift=10, yshift=-9, color=YellowOrange]$\vv{N^-}$}]();
		\pic["$\alpha^-$", color=BrickRed, draw=BrickRed, fill=BrickRed!20, ->, angle eccentricity= 1.5, xshift=100]{angle = No--Pw--Do};
		\draw[->, line width=1] (Pw)->(To);
		\draw[->, line width=1, color=black!60!green] (Pw)->(Do);
		\draw[->, line width=1] (Pw)->(No);
		\draw[dashed, color=lightgray] (Pw)--($(Pw)-10*(dir)$);
	}
	
	{
		\pgfmathparse{cos(360/\sizeTex*\point)}
		\pgfmathsetmacro {\PvalueX}{\pgfmathresult}
		\pgfmathparse{sin(360/\sizeTex*\point)}
		\pgfmathsetmacro {\PvalueY}{\pgfmathresult}
		\coordinate (Pw) at ($10*(\PvalueX,\PvalueY)$);
		\coordinate[label={[xshift=-6, yshift=-10]$\hat{e}_z$}] (No) at ($11*(\PvalueX,\PvalueY)$);
		
		\pgfmathparse{cos(360/\sizeTex*\point+90)}
		\pgfmathsetmacro {\TvalueX}{\pgfmathresult}
		\pgfmathparse{sin(360/\sizeTex*\point+90)}
		\pgfmathsetmacro {\TvalueY}{\pgfmathresult}
		\coordinate[label={[xshift=3, yshift=-15]$\hat{e}_x$}] (To) at ($(Pw)+(\TvalueX,\TvalueY)$);
		
		\coordinate[label={[xshift=4, yshift=-9, color=black!60!green]$\hat{d}$}] (Do) at ($(Pw)+(dir)$);
		
		\draw[->, line width=1, color=YellowOrange] (Pw)->($2*(No)-(Pw)$) coordinate[label={[xshift=10, yshift=-9, color=YellowOrange]$\vv{N}$}]();
		\pic["$\alpha$", color=BrickRed, draw=BrickRed, fill=BrickRed!20, ->, angle eccentricity= 1.5, xshift=100]{angle = No--Pw--Do};
		\draw[->, line width=1] (Pw)->(To);
		\draw[->, line width=1, color=black!60!green] (Pw)->(Do);
		\draw[->, line width=1] (Pw)->(No);
		\draw[dashed, color=lightgray] (Pw)--($(Pw)-10*(dir)$);
	}
	
	{
		\pgfmathparse{cos(360/\sizeTex*\pointplus)}
		\pgfmathsetmacro {\PvalueX}{\pgfmathresult}
		\pgfmathparse{sin(360/\sizeTex*\pointplus)}
		\pgfmathsetmacro {\PvalueY}{\pgfmathresult}
		\coordinate[label={[xshift=5, yshift=0.3]$P^+$}] (Pw) at ($10*(\PvalueX,\PvalueY)$);
		\coordinate[label={[xshift=-5, yshift=-13]$\hat{e}^+_z$}] (No) at ($11*(\PvalueX,\PvalueY)$);
		
		\pgfmathparse{cos(360/\sizeTex*\pointplus+90)}
		\pgfmathsetmacro {\TvalueX}{\pgfmathresult}
		\pgfmathparse{sin(360/\sizeTex*\pointplus+90)}
		\pgfmathsetmacro {\TvalueY}{\pgfmathresult}
		\coordinate[label={[xshift=7, yshift=-16]$\hat{e}^+_x$}] (To) at ($(Pw)+(\TvalueX,\TvalueY)$);
		
		\coordinate[label={[xshift=4, yshift=-9, color=black!60!green]$\hat{d}$}] (Do) at ($(Pw)+(dir)$);
		
		\draw[->, line width=1, color=YellowOrange] (Pw)->($2*(No)-(Pw)$) coordinate[label={[xshift=10, yshift=-9, color=YellowOrange]$\vv{N^+}$}]();
		\pic[pic text=$\alpha^+$,pic text options={xshift = 6, yshift=15}, color=BrickRed, draw=BrickRed, fill=BrickRed!20, ->, angle eccentricity= 1.5, xshift=100]{angle = No--Pw--Do};
		\draw[->, line width=1] (Pw)->(To);
		\draw[->, line width=1, color=black!60!green] (Pw)->(Do);
		\draw[->, line width=1] (Pw)->(No);
		\draw[dashed, color=lightgray] (Pw)--($(Pw)-10*(dir)$);
	}
	
\end{tikzpicture}
		}
	}
	\centering
	\caption{2D representation of the render process.}
	\label{fig:render2D}
\end{figure}

 Fig. \ref{fig:render2D} illustrates this idea using a 2D representation, note that the two points ($P^x$) illustrated in the two dimensional simplification are actually four points in a three dimensional model and represents the closest points to $P$ of our surface that matches a pixel in our discrete texture. The observer $O$ views the point $P$ over the surface $S$, the colour displayed by $P$ may be calculated by:

\begin{enumerate}
  \item Calculate the direction view ($\vv{d}$);
  \item find the four closest points over our surface that best match our discrete texture ($P^x$);
  \item calculate the angles ($\alpha	^x$) for each point ($P^x$) using the direction view ($\vv{d}$), considering their local coordinate system;
	\item convert the values of position and angle of the four points to $u$, $v$, $s$, and $t$ using (\ref{eq:sphere_par})(for the sphere case), (\ref{eq:second_coord_system}) and (\ref{eq:angula_dimension});
	\item use the combinations of $u$, $v$, $s$, and $t$ to acquire 16 colour values from our texture;
	\item apply bilinear interpolation over the angular dimensions, achieving four colour values for the four points ($P^x$), note that each point has different interpolation’s weight for $s$ and $t$;
	\item apply bilinear interpolation on the spatial dimensions to reduce the four values to one final pixel value.
\end{enumerate}

\section{Texture Restrictions}

\begin{figure}
	\centering
	\begin{subfigure}{.3\textwidth}
		\centering
		\fbox{
			\resizebox{.8\textwidth}{!}{
				\includegraphics{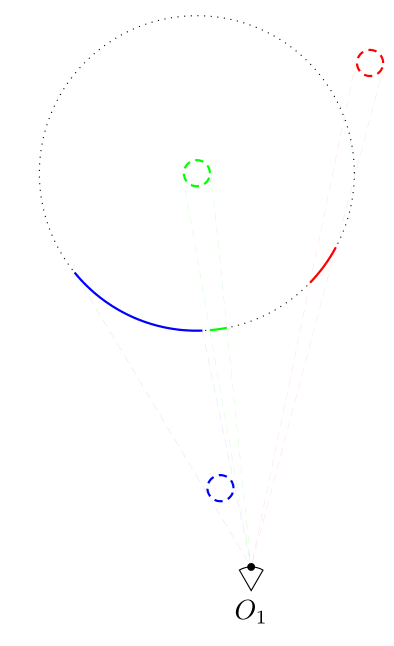}
			}
		}
		\caption{}
		\label{fig:rest1}
	\end{subfigure}%
	\begin{subfigure}{.3\textwidth}
		\centering
		\fbox{
			\resizebox{.8\textwidth}{!}{
				\includegraphics{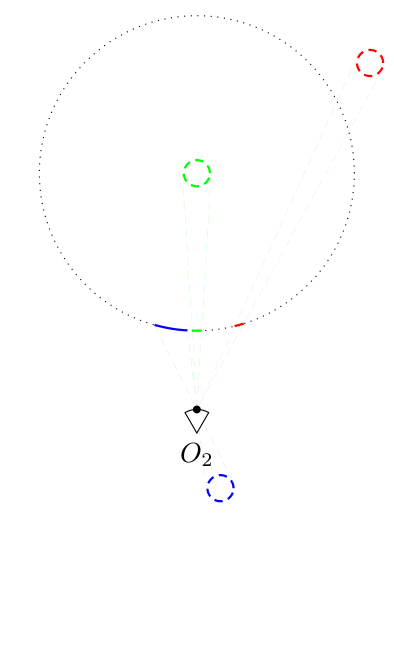}
			}
		}
		\caption{}
		\label{fig:rest2}
	\end{subfigure}
	\begin{subfigure}{.3\textwidth}
		\centering
		\fbox{
			\resizebox{.8\textwidth}{!}{
				\includegraphics{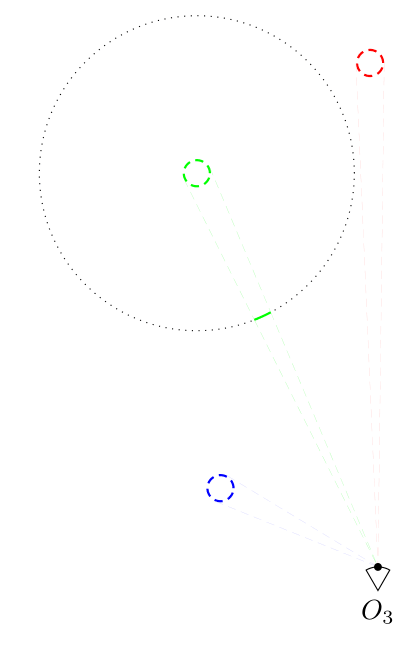}
			}
		}
		\caption{}
		\label{fig:rest3}
	\end{subfigure}
	\caption{2D Illustration showing the effect of differenties observer's placements.}
	\label{fig:rest}
\end{figure}

An object or scene is able to be represented as a simplified model as long it is able to display an illusion of the real objects for an observer in the desired position. In section \ref{sec:introduction} we mentioned that this texture was thought to be able to represent positive parallax of any real surface inside our model object, Fig. \ref{fig:rest} presents a 2D representation to illustrate these limitations.

On Fig. \ref{fig:rest} the red, green and blue circles represents three objects and the dotted black circle the simplified model that will be used to represent our three objects. The three sub-images presents three different conditions of placement for our viewer $O$. Our viewer is not able to see directly the three objects, only the solid lines compose the image viewable from its point of view.
\par On Fig. \ref{fig:rest1} the observer is placed behind all objects, in a way that the projection of all coloured objects superpose the model, and therefore he is able to correctly see the blue object with negative parallax and the green and red objects with positive parallax.
\par On Fig. \ref{fig:rest2} we placed the observer between the blue object and the model, and he is still able to see the green and red objects correctly, but the blue object incorrectly appears as an projection of itself over our surface model.
\par At last on Fig. \ref{fig:rest3} the observer is placed in a way that the projection of the blue and red objects does not superpose the model, this way there are no projections of this two objects over our surface and therefore our viewer is not able to see them. About the green object, it is seen correctly as usual.
\par From the three objects, only the green one, that is inside our model and is always seen through positive parallax, is seen correctly in the three studied cases and therefore allows the placement of an observer outside the model with no restriction. For objects positioned outside the model we add the condition that the projection of the object over the observer's position must meet the simplified model, and for negative parallax that the intended object must be between the observer and the simplified model's surface.

\section{Test Scene}

Two virtual test scenes were created. The first one was a scene composed by only one primitive object with the objective of investigate the effect of texture resolution on aliasing(\ref{sec:aliasscene}). The second was created by disposing several standard primitive objects in different places and angles to check the effect of different resolutions over composed scenes(\ref{sec:composedscene}).

\subsection{Aliasing Scene}
\label{sec:aliasscene}
A scene with a unique moving cylinder of diameter 1 and height 2 was created with the objective to visualize the effect of aliasing with the increment of the distance between the simplified model surface and the object. Two positions were selected to represent the desired effect.
\par A texture of 1024x512x32x32 was generated using a spheric model of diameter 7 through a ray-tracing process starting from the models surface, pointed to its inside. The points sampled were all the ones that match our texture parametrization over our surface model. The angular dimensions were purposely kept low to emphasize the desired effect over this dimensions.
\par
\begin{figure}[h]
  \centering
  \begin{tabular}{|m{3cm}|m{3cm}|}
		\hline
		Scene & Image \\
		\hline
		\resizebox{\linewidth}{!}{
			\pgfmathsetmacro{\posx}{5}
			\pgfmathsetmacro{\posy}{0}
			\pgfmathsetmacro{\posz}{0}
			\pgfmathsetmacro{\dirx}{-2}
			\pgfmathsetmacro{\diry}{0}
			\pgfmathsetmacro{\dirz}{0}
			\pgfmathsetmacro{\cilx}{2}
			\pgfmathsetmacro{\cily}{0}
			\pgfmathsetmacro{\cilz}{0}
			\tdplotsetmaincoords{60}{110}
	\begin{tikzpicture}[tdplot_main_coords, scale = .3]

		\newcommand{\tdsphere}[5]{%
			\shade[ball color = #5,
					opacity = 1
			] (#3,#1,#2) circle (#4cm);
		}
		
		\newcommand{\tdcylinderVertical}[6]{%
			\begin{scope}[x={(1,0)}]
				\path (1,0,0);
				\pgfgetlastxy{\cylxx}{\cylxy}
				\path (0,1,0);
				\pgfgetlastxy{\cylyx}{\cylyy}
				\path (0,0,1);
				\pgfgetlastxy{\cylzx}{\cylzy}
				\pgfmathsetmacro{\cylt}{(\cylzy * \cylyx - \cylzx * \cylyy)/ (\cylzy * \cylxx - \cylzx * \cylxy)}
				\pgfmathsetmacro{\ang}{atan(\cylt)}
				\pgfmathsetmacro{\ct}{1/sqrt(1 + (\cylt)^2)}
				\pgfmathsetmacro{\st}{\cylt * \ct}
				\shade[ball color = #6, opacity = 1] (#3+#4*\ct,#1+#4*\st,#2+#5/2) -- ++(0,0,-1*#5) arc[start angle=\ang,delta angle=180,radius=#4] -- ++(0,0,#5) arc[start angle=\ang+180,delta angle=-180,radius=#4];
				\begin{scope}[every path/.style={ultra thick}]
					\shade[ball color = #6, opacity = 1] (#3,#1,#2+#5/2) circle[radius=#4];
				\end{scope}
			\end{scope}
		}
		
		\newcommand{\tdcylinderHorizontal}[6]{%
			\begin{scope}[x={(1,0)}]
				\path (1,0,0);
				\pgfgetlastxy{\cylxx}{\cylxy}
				\path (0,1,0);
				\pgfgetlastxy{\cylyx}{\cylyy}
				\path (0,0,1);
				\pgfgetlastxy{\cylzx}{\cylzy}
				\pgfmathsetmacro{\cylt}{(\cylxy * \cylyx - \cylxx * \cylyy)/ (\cylxy * \cylzx - \cylxx * \cylzy)}
				\pgfmathsetmacro{\ang}{atan(\cylt)}
				\pgfmathsetmacro{\ct}{1/sqrt(1 + (\cylt)^2)}
				\pgfmathsetmacro{\st}{\cylt * \ct}
				
				\shade[ball color = #6, opacity = 1] (#3+#5/2,#1-#4*\st,#2-#4*\ct) -- ++(-#5,0,0) {[canvas is yz plane at x=#3-#5/2] arc[start angle=\ang+30,delta angle=180,radius=#4]} -- ++(#5,0,0) {[canvas is yz plane at x=#3+#5/2] arc[start angle=\ang+180+30,delta angle=-180,radius=#4]};
				\begin{scope}[every path/.style={ultra thick}, canvas is yz plane at x=#3+#5/2	]
					\shade[ball color = #6, opacity = 1] (#1,#2,0) circle[radius=#4];
				\end{scope}
			\end{scope}
		}
		
		\newcommand{\tdcube}[7]{%
				\shade[ball color = #7, opacity = 1] (#3-#6/2,#1-#4/2,#2+#5/2) -- ++(#6,0,0) -- ++(0,#4,0) -- ++(-#6,0,0) -- ++ (0,-#4,0);
				\shade[ball color = #7, opacity = 1] (#3+#6/2,#1-#4/2,#2+#5/2) -- ++(0,#4,0) -- ++(0,0,-#5) -- ++(0,-#4,0) -- ++ (0,0,#5);
				\shade[ball color = #7, opacity = 1] (#3+#6/2,#1+#4/2,#2+#5/2) -- ++(-#6,0,0) -- ++(0,0,-#5) -- ++(#6,0,0) -- ++ (0,0,#5);
		}
	 
		\draw[-stealth] (0,0,0) -- (11,0,0) 
				node[below left] {\huge $z$};
		 
		\draw[-stealth] (0,0,0) -- (0,11,0)
				node[below right] {\huge $x$};
		 
		\draw[-stealth] (0,0,0) -- (0,0,11)
				node[above] {\huge $y$};
				
		\draw[dashed] (0,0,0) -- (-11,0,0);
		 
		\draw[dashed] (0,0,0) -- (0,-11,0);
		 
		\draw[dashed] (0,0,0) -- (0,0,-11);

		\coordinate (P) at (\posz,\posx,\posy);
		\coordinate (D) at (\dirz,\dirx,\diry);
		\draw[red,fill=red] (P) circle (.5ex);
		\draw[-{stealth[scale=2]},color=red, very thick] (P) -- ($(P)+(D)$);
		
		\tdcylinderVertical{\cilx}{\cily}{\cilz}{.5}{2}{violet}
	 
	\end{tikzpicture}
		} & \includegraphics[height=\linewidth]{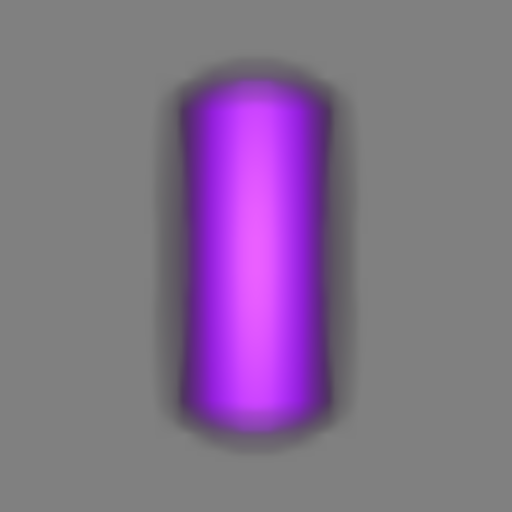} \\
		\hline
    \resizebox{\linewidth}{!}{
			\pgfmathsetmacro{\posx}{5}
			\pgfmathsetmacro{\posy}{0}
			\pgfmathsetmacro{\posz}{0}
			\pgfmathsetmacro{\dirx}{-2}
			\pgfmathsetmacro{\diry}{0}
			\pgfmathsetmacro{\dirz}{0}
			\pgfmathsetmacro{\cilx}{-5}
			\pgfmathsetmacro{\cily}{0}
			\pgfmathsetmacro{\cilz}{0}
			\tdplotsetmaincoords{60}{110}
	\begin{tikzpicture}[tdplot_main_coords, scale = .3]

		\newcommand{\tdsphere}[5]{%
			\shade[ball color = #5,
					opacity = 1
			] (#3,#1,#2) circle (#4cm);
		}
		
		\newcommand{\tdcylinderVertical}[6]{%
			\begin{scope}[x={(1,0)}]
				\path (1,0,0);
				\pgfgetlastxy{\cylxx}{\cylxy}
				\path (0,1,0);
				\pgfgetlastxy{\cylyx}{\cylyy}
				\path (0,0,1);
				\pgfgetlastxy{\cylzx}{\cylzy}
				\pgfmathsetmacro{\cylt}{(\cylzy * \cylyx - \cylzx * \cylyy)/ (\cylzy * \cylxx - \cylzx * \cylxy)}
				\pgfmathsetmacro{\ang}{atan(\cylt)}
				\pgfmathsetmacro{\ct}{1/sqrt(1 + (\cylt)^2)}
				\pgfmathsetmacro{\st}{\cylt * \ct}
				\shade[ball color = #6, opacity = 1] (#3+#4*\ct,#1+#4*\st,#2+#5/2) -- ++(0,0,-1*#5) arc[start angle=\ang,delta angle=180,radius=#4] -- ++(0,0,#5) arc[start angle=\ang+180,delta angle=-180,radius=#4];
				\begin{scope}[every path/.style={ultra thick}]
					\shade[ball color = #6, opacity = 1] (#3,#1,#2+#5/2) circle[radius=#4];
				\end{scope}
			\end{scope}
		}
		
		\newcommand{\tdcylinderHorizontal}[6]{%
			\begin{scope}[x={(1,0)}]
				\path (1,0,0);
				\pgfgetlastxy{\cylxx}{\cylxy}
				\path (0,1,0);
				\pgfgetlastxy{\cylyx}{\cylyy}
				\path (0,0,1);
				\pgfgetlastxy{\cylzx}{\cylzy}
				\pgfmathsetmacro{\cylt}{(\cylxy * \cylyx - \cylxx * \cylyy)/ (\cylxy * \cylzx - \cylxx * \cylzy)}
				\pgfmathsetmacro{\ang}{atan(\cylt)}
				\pgfmathsetmacro{\ct}{1/sqrt(1 + (\cylt)^2)}
				\pgfmathsetmacro{\st}{\cylt * \ct}
				
				\shade[ball color = #6, opacity = 1] (#3+#5/2,#1-#4*\st,#2-#4*\ct) -- ++(-#5,0,0) {[canvas is yz plane at x=#3-#5/2] arc[start angle=\ang+30,delta angle=180,radius=#4]} -- ++(#5,0,0) {[canvas is yz plane at x=#3+#5/2] arc[start angle=\ang+180+30,delta angle=-180,radius=#4]};
				\begin{scope}[every path/.style={ultra thick}, canvas is yz plane at x=#3+#5/2	]
					\shade[ball color = #6, opacity = 1] (#1,#2,0) circle[radius=#4];
				\end{scope}
			\end{scope}
		}
		
		\newcommand{\tdcube}[7]{%
				\shade[ball color = #7, opacity = 1] (#3-#6/2,#1-#4/2,#2+#5/2) -- ++(#6,0,0) -- ++(0,#4,0) -- ++(-#6,0,0) -- ++ (0,-#4,0);
				\shade[ball color = #7, opacity = 1] (#3+#6/2,#1-#4/2,#2+#5/2) -- ++(0,#4,0) -- ++(0,0,-#5) -- ++(0,-#4,0) -- ++ (0,0,#5);
				\shade[ball color = #7, opacity = 1] (#3+#6/2,#1+#4/2,#2+#5/2) -- ++(-#6,0,0) -- ++(0,0,-#5) -- ++(#6,0,0) -- ++ (0,0,#5);
		}
	 
		\draw[-stealth] (0,0,0) -- (11,0,0) 
				node[below left] {\huge $z$};
		 
		\draw[-stealth] (0,0,0) -- (0,11,0)
				node[below right] {\huge $x$};
		 
		\draw[-stealth] (0,0,0) -- (0,0,11)
				node[above] {\huge $y$};
				
		\draw[dashed] (0,0,0) -- (-11,0,0);
		 
		\draw[dashed] (0,0,0) -- (0,-11,0);
		 
		\draw[dashed] (0,0,0) -- (0,0,-11);

		\coordinate (P) at (\posz,\posx,\posy);
		\coordinate (D) at (\dirz,\dirx,\diry);
		\draw[red,fill=red] (P) circle (.5ex);
		\draw[-{stealth[scale=2]},color=red, very thick] (P) -- ($(P)+(D)$);
		
		\tdcylinderVertical{\cilx}{\cily}{\cilz}{.5}{2}{violet}
	 
	\end{tikzpicture}
		} & \includegraphics[height=\linewidth]{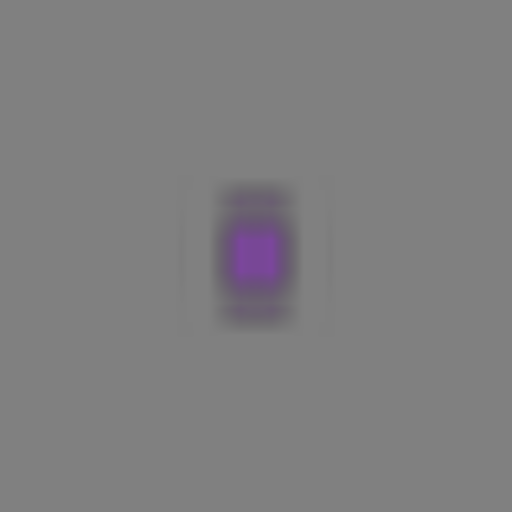} \\
		\hline
  \end{tabular}
  \caption{Images showing the different aliasing effect produced by moving an object away from the model surface.}
  \label{tab:comparative_aliasing}
\end{figure}
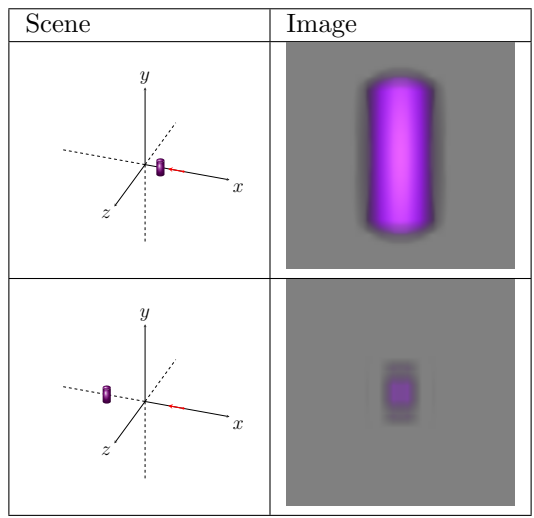

 As seen on fig. \ref{tab:comparative_aliasing} the projection of the cylinder over our model becomes blurrier as the object gets away from the viewable surface model. After a certain distance an effect of discontinuity appears because of our render algorithm.

\subsection{Composed Scene}
\label{sec:composedscene}
This scene is composed by 2 spheres, 3 cylinders and 2 cubes. Both spheres has diameter 1, the cubes were regular of side 1 and the cylinders diameter 1 and height 2. Table \ref{table:list_object} shows the position and rotation of each object in this scene. Also one directional light pointed to the direction $[-1,0,0]^T$ was added to our scene to produce effects of diffuse and specular reflection.

\begin{table}[ht]
	\centering
	\resizebox{.45\textwidth}{!}{
	\begin{tabular}{|l|c|c|c|c|c|c|}
		\hline
		\multirow{2}{*}{Colour} & \multicolumn{3}{|c|}{Position} & \multicolumn{3}{|c|}{Rotation} \\
		\cline{2-7}
		& x & y & z & x & y & z\\
		\hline
		\multicolumn{7}{|c|}{Sphere}\\
		\hline
		White & 0 & 0 & 0 & 0 & 0 & 0\\
		Red & 0 & 0 & -1.87 & 0 & 0 & 0\\
		\hline
		\multicolumn{7}{|c|}{Cylinder}\\
		\hline
		Purple & 2.62 & 0 & 0 & 0 &0 &0\\
		Blue & -1.07 & -0.41 & 1.81 & 90$^{\circ}$ & 0 & 0\\
		Black & 0.87 & 0.76 & 2.06 & 0 & 0 & 0\\
		\hline
		\multicolumn{7}{|c|}{Cube}\\
		\hline
		Green & -1.2 & 0.71 & -0.68 & 0 & 0 & 0\\
		Yellow & 1.47 & -1.05 & 0.93 & 0 & 0 & 0\\
		\hline
	\end{tabular}
	}
	\caption{List of objects on the Composed Scene.}
	\label{table:list_object}
\end{table} 

\begin{figure}
	\centering
	\begin{subfigure}{.23\textwidth}
		\centering
		\includegraphics[width=.8\linewidth]{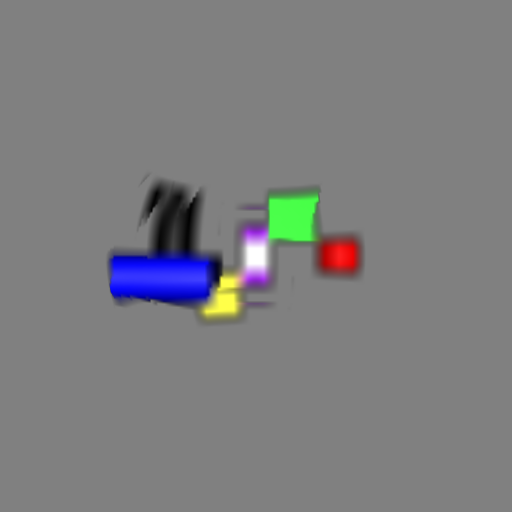}
		\caption{}
		\label{fig:aliasTex}
	\end{subfigure}%
	\begin{subfigure}{.23\textwidth}
		\centering
		\includegraphics[width=.8\linewidth]{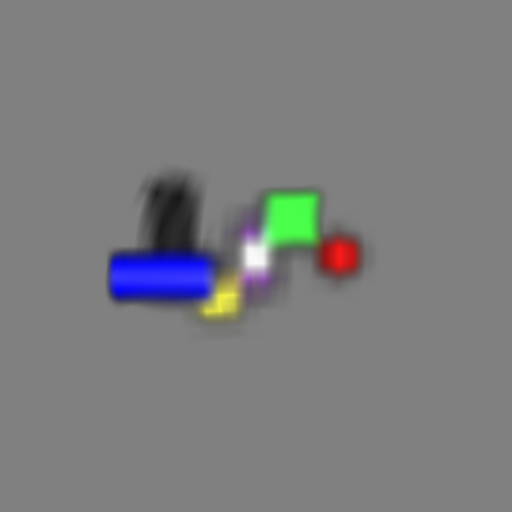}
		\caption{}
		\label{fig:antiAliasTex}
	\end{subfigure}
	\caption{Comparison between textures without(\ref{fig:aliasTex}) and with(\ref{fig:antiAliasTex}) 7x supersample anti-aliasing method}
	\label{fig:compareAlias}
\end{figure}

This scene was approximated to a sphere of diameter 7 centred at the origin. Our texture was also generated through a ray tracing process starting from our simplified model surface, the points sampled were all the ones that match our texture parametrization. A simple direct sampling produced an undesirable aliasing effect (Fig. \ref{fig:aliasTex}). To minimize this effect and obtain a better result, a 7x supersample anti-aliasing method was used over all dimensions during the texture synthesis(Fig. \ref{fig:antiAliasTex}).

\begin{figure*}[t]
  \centering
  \begin{tabular}{|m{2.9cm}|m{2cm}|m{2cm}|m{2cm}|m{2cm}|m{2cm}|}
		\hline
		\small{Position and Direction} & \small{Direct Render} & \small{U:1024 V:512 S:128 T:128} & \small{U:512 V:256 S:32 T:32} & \small{U:64 V:32 S:256 T:256} & \small{U:512 V:256 S:64 T:64} \\
		\hline
		\resizebox{\linewidth}{!}{
			\pgfmathsetmacro{\posx}{10}
			\pgfmathsetmacro{\posy}{0}
			\pgfmathsetmacro{\posz}{0}
			\pgfmathsetmacro{\dirx}{-3}
			\pgfmathsetmacro{\diry}{0}
			\pgfmathsetmacro{\dirz}{0}
			\tdplotsetmaincoords{60}{110}
	\begin{tikzpicture}[tdplot_main_coords, scale = .3]

		\newcommand{\tdsphere}[5]{%
			\shade[ball color = #5,
					opacity = 1
			] (#3,#1,#2) circle (#4cm);
		}
		
		\newcommand{\tdcylinderVertical}[6]{%
			\begin{scope}[x={(1,0)}]
				\path (1,0,0);
				\pgfgetlastxy{\cylxx}{\cylxy}
				\path (0,1,0);
				\pgfgetlastxy{\cylyx}{\cylyy}
				\path (0,0,1);
				\pgfgetlastxy{\cylzx}{\cylzy}
				\pgfmathsetmacro{\cylt}{(\cylzy * \cylyx - \cylzx * \cylyy)/ (\cylzy * \cylxx - \cylzx * \cylxy)}
				\pgfmathsetmacro{\ang}{atan(\cylt)}
				\pgfmathsetmacro{\ct}{1/sqrt(1 + (\cylt)^2)}
				\pgfmathsetmacro{\st}{\cylt * \ct}
				\shade[ball color = #6, opacity = 1] (#3+#4*\ct,#1+#4*\st,#2+#5/2) -- ++(0,0,-1*#5) arc[start angle=\ang,delta angle=180,radius=#4] -- ++(0,0,#5) arc[start angle=\ang+180,delta angle=-180,radius=#4];
				\begin{scope}[every path/.style={ultra thick}]
					\shade[ball color = #6, opacity = 1] (#3,#1,#2+#5/2) circle[radius=#4];
				\end{scope}
			\end{scope}
		}
		
		\newcommand{\tdcylinderHorizontal}[6]{%
			\begin{scope}[x={(1,0)}]
				\path (1,0,0);
				\pgfgetlastxy{\cylxx}{\cylxy}
				\path (0,1,0);
				\pgfgetlastxy{\cylyx}{\cylyy}
				\path (0,0,1);
				\pgfgetlastxy{\cylzx}{\cylzy}
				\pgfmathsetmacro{\cylt}{(\cylxy * \cylyx - \cylxx * \cylyy)/ (\cylxy * \cylzx - \cylxx * \cylzy)}
				\pgfmathsetmacro{\ang}{atan(\cylt)}
				\pgfmathsetmacro{\ct}{1/sqrt(1 + (\cylt)^2)}
				\pgfmathsetmacro{\st}{\cylt * \ct}
				
				\shade[ball color = #6, opacity = 1] (#3+#5/2,#1-#4*\st,#2-#4*\ct) -- ++(-#5,0,0) {[canvas is yz plane at x=#3-#5/2] arc[start angle=\ang+30,delta angle=180,radius=#4]} -- ++(#5,0,0) {[canvas is yz plane at x=#3+#5/2] arc[start angle=\ang+180+30,delta angle=-180,radius=#4]};
				\begin{scope}[every path/.style={ultra thick}, canvas is yz plane at x=#3+#5/2	]
					\shade[ball color = #6, opacity = 1] (#1,#2,0) circle[radius=#4];
				\end{scope}
			\end{scope}
		}
		
		\newcommand{\tdcube}[7]{%
				\shade[ball color = #7, opacity = 1] (#3-#6/2,#1-#4/2,#2+#5/2) -- ++(#6,0,0) -- ++(0,#4,0) -- ++(-#6,0,0) -- ++ (0,-#4,0);
				\shade[ball color = #7, opacity = 1] (#3+#6/2,#1-#4/2,#2+#5/2) -- ++(0,#4,0) -- ++(0,0,-#5) -- ++(0,-#4,0) -- ++ (0,0,#5);
				\shade[ball color = #7, opacity = 1] (#3+#6/2,#1+#4/2,#2+#5/2) -- ++(-#6,0,0) -- ++(0,0,-#5) -- ++(#6,0,0) -- ++ (0,0,#5);
		}
	 
		\draw[-stealth] (0,0,0) -- (11,0,0) 
				node[below left] {\huge $z$};
		 
		\draw[-stealth] (0,0,0) -- (0,11,0)
				node[below right] {\huge $x$};
		 
		\draw[-stealth] (0,0,0) -- (0,0,11)
				node[above] {\huge $y$};
				
		\draw[dashed] (0,0,0) -- (-11,0,0);
		 
		\draw[dashed] (0,0,0) -- (0,-11,0);
		 
		\draw[dashed] (0,0,0) -- (0,0,-11);

		\coordinate (P) at (\posz,\posx,\posy);
		\coordinate (D) at (\dirz,\dirx,\diry);
		\draw[red,fill=red] (P) circle (.5ex);
		\draw[-{stealth[scale=2]},color=red, very thick] (P) -- ($(P)+(D)$);
		
		\tdsphere{0}{0}{0}{.5}{white}
		\tdsphere{0}{0}{-1.87}{.5}{red}
		\tdcylinderVertical{-2.62}{0}{0}{.5}{2}{violet}
		\tdcylinderHorizontal{1.07}{-.41}{1.81}{.5}{2}{blue}
		\tdcube{-1.47}{-1.05}{0.93}{1}{1}{1}{yellow}
		\tdcylinderVertical{-0.87}{0.76}{2.06}{.5}{2}{black}
		\tdcube{1.2}{0.71}{-0.68}{1}{1}{1}{green}
	 
	\end{tikzpicture}
		} & \includegraphics[height=\linewidth]{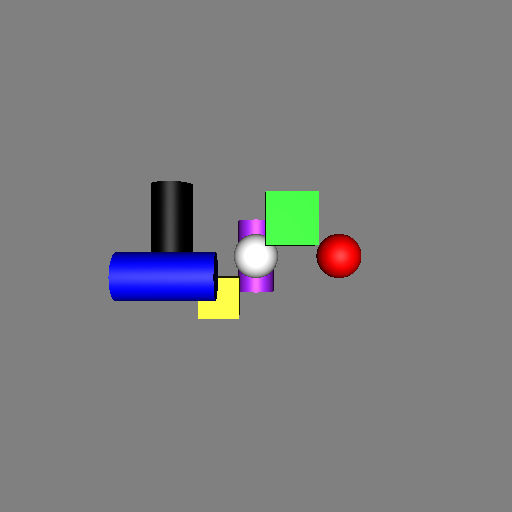} & \includegraphics[height=\linewidth]{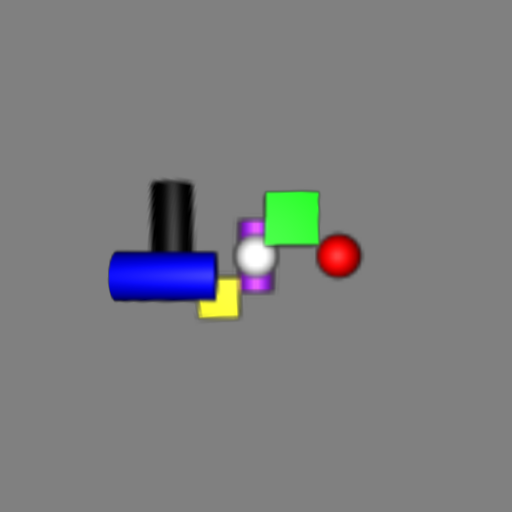} & \includegraphics[width=\linewidth]{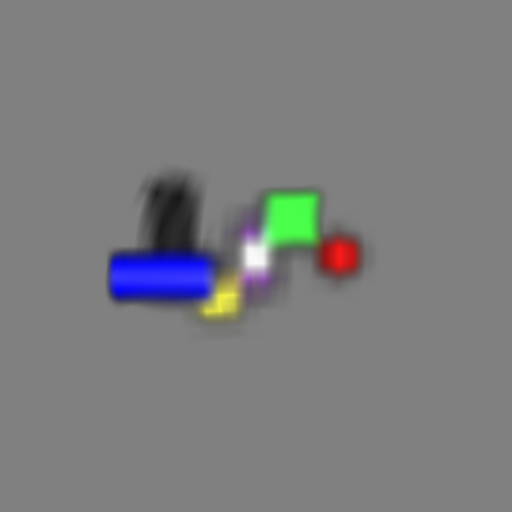}& \includegraphics[width=\linewidth]{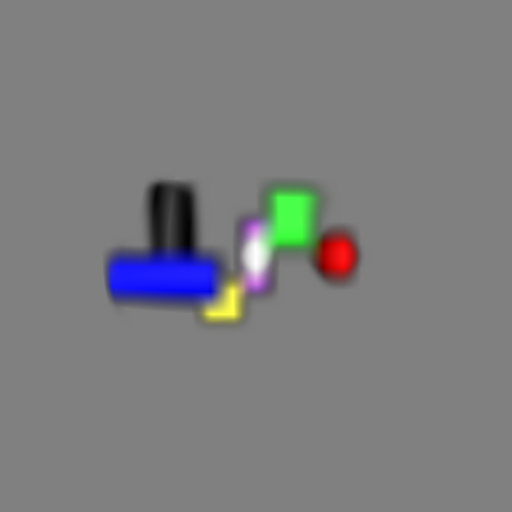} & \includegraphics[width=\linewidth]{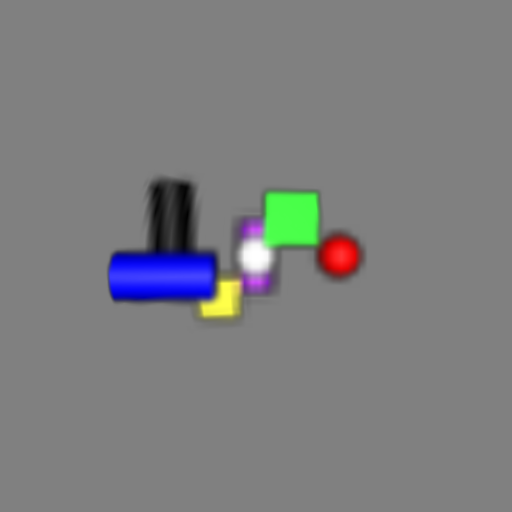} \\
		\hline
    \resizebox{\linewidth}{!}{
			\pgfmathsetmacro{\posx}{10}
			\pgfmathsetmacro{\posy}{0}
			\pgfmathsetmacro{\posz}{-3}
			\pgfmathsetmacro{\dirx}{-3}
			\pgfmathsetmacro{\diry}{0}
			\pgfmathsetmacro{\dirz}{0}
			\tdplotsetmaincoords{60}{110}
	\begin{tikzpicture}[tdplot_main_coords, scale = .3]

		\newcommand{\tdsphere}[5]{%
			\shade[ball color = #5,
					opacity = 1
			] (#3,#1,#2) circle (#4cm);
		}
		
		\newcommand{\tdcylinderVertical}[6]{%
			\begin{scope}[x={(1,0)}]
				\path (1,0,0);
				\pgfgetlastxy{\cylxx}{\cylxy}
				\path (0,1,0);
				\pgfgetlastxy{\cylyx}{\cylyy}
				\path (0,0,1);
				\pgfgetlastxy{\cylzx}{\cylzy}
				\pgfmathsetmacro{\cylt}{(\cylzy * \cylyx - \cylzx * \cylyy)/ (\cylzy * \cylxx - \cylzx * \cylxy)}
				\pgfmathsetmacro{\ang}{atan(\cylt)}
				\pgfmathsetmacro{\ct}{1/sqrt(1 + (\cylt)^2)}
				\pgfmathsetmacro{\st}{\cylt * \ct}
				\shade[ball color = #6, opacity = 1] (#3+#4*\ct,#1+#4*\st,#2+#5/2) -- ++(0,0,-1*#5) arc[start angle=\ang,delta angle=180,radius=#4] -- ++(0,0,#5) arc[start angle=\ang+180,delta angle=-180,radius=#4];
				\begin{scope}[every path/.style={ultra thick}]
					\shade[ball color = #6, opacity = 1] (#3,#1,#2+#5/2) circle[radius=#4];
				\end{scope}
			\end{scope}
		}
		
		\newcommand{\tdcylinderHorizontal}[6]{%
			\begin{scope}[x={(1,0)}]
				\path (1,0,0);
				\pgfgetlastxy{\cylxx}{\cylxy}
				\path (0,1,0);
				\pgfgetlastxy{\cylyx}{\cylyy}
				\path (0,0,1);
				\pgfgetlastxy{\cylzx}{\cylzy}
				\pgfmathsetmacro{\cylt}{(\cylxy * \cylyx - \cylxx * \cylyy)/ (\cylxy * \cylzx - \cylxx * \cylzy)}
				\pgfmathsetmacro{\ang}{atan(\cylt)}
				\pgfmathsetmacro{\ct}{1/sqrt(1 + (\cylt)^2)}
				\pgfmathsetmacro{\st}{\cylt * \ct}
				
				\shade[ball color = #6, opacity = 1] (#3+#5/2,#1-#4*\st,#2-#4*\ct) -- ++(-#5,0,0) {[canvas is yz plane at x=#3-#5/2] arc[start angle=\ang+30,delta angle=180,radius=#4]} -- ++(#5,0,0) {[canvas is yz plane at x=#3+#5/2] arc[start angle=\ang+180+30,delta angle=-180,radius=#4]};
				\begin{scope}[every path/.style={ultra thick}, canvas is yz plane at x=#3+#5/2	]
					\shade[ball color = #6, opacity = 1] (#1,#2,0) circle[radius=#4];
				\end{scope}
			\end{scope}
		}
		
		\newcommand{\tdcube}[7]{%
				\shade[ball color = #7, opacity = 1] (#3-#6/2,#1-#4/2,#2+#5/2) -- ++(#6,0,0) -- ++(0,#4,0) -- ++(-#6,0,0) -- ++ (0,-#4,0);
				\shade[ball color = #7, opacity = 1] (#3+#6/2,#1-#4/2,#2+#5/2) -- ++(0,#4,0) -- ++(0,0,-#5) -- ++(0,-#4,0) -- ++ (0,0,#5);
				\shade[ball color = #7, opacity = 1] (#3+#6/2,#1+#4/2,#2+#5/2) -- ++(-#6,0,0) -- ++(0,0,-#5) -- ++(#6,0,0) -- ++ (0,0,#5);
		}
	 
		\draw[-stealth] (0,0,0) -- (11,0,0) 
				node[below left] {\huge $z$};
		 
		\draw[-stealth] (0,0,0) -- (0,11,0)
				node[below right] {\huge $x$};
		 
		\draw[-stealth] (0,0,0) -- (0,0,11)
				node[above] {\huge $y$};
				
		\draw[dashed] (0,0,0) -- (-11,0,0);
		 
		\draw[dashed] (0,0,0) -- (0,-11,0);
		 
		\draw[dashed] (0,0,0) -- (0,0,-11);

		\coordinate (P) at (\posz,\posx,\posy);
		\coordinate (D) at (\dirz,\dirx,\diry);
		\draw[red,fill=red] (P) circle (.5ex);
		\draw[-{stealth[scale=2]},color=red, very thick] (P) -- ($(P)+(D)$);
		
		\tdsphere{0}{0}{0}{.5}{white}
		\tdsphere{0}{0}{-1.87}{.5}{red}
		\tdcylinderVertical{-2.62}{0}{0}{.5}{2}{violet}
		\tdcylinderHorizontal{1.07}{-.41}{1.81}{.5}{2}{blue}
		\tdcube{-1.47}{-1.05}{0.93}{1}{1}{1}{yellow}
		\tdcylinderVertical{-0.87}{0.76}{2.06}{.5}{2}{black}
		\tdcube{1.2}{0.71}{-0.68}{1}{1}{1}{green}
	 
	\end{tikzpicture}
		} & \includegraphics[height=\linewidth]{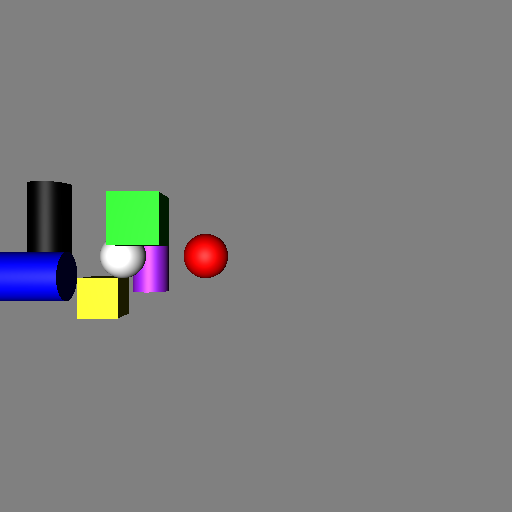} & \includegraphics[width=\linewidth]{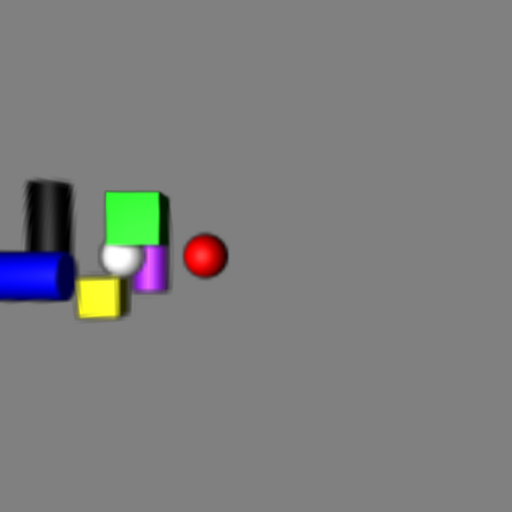} & \includegraphics[width=\linewidth]{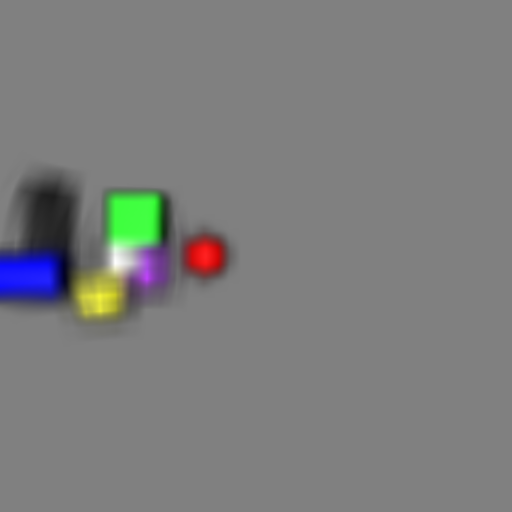}& \includegraphics[width=\linewidth]{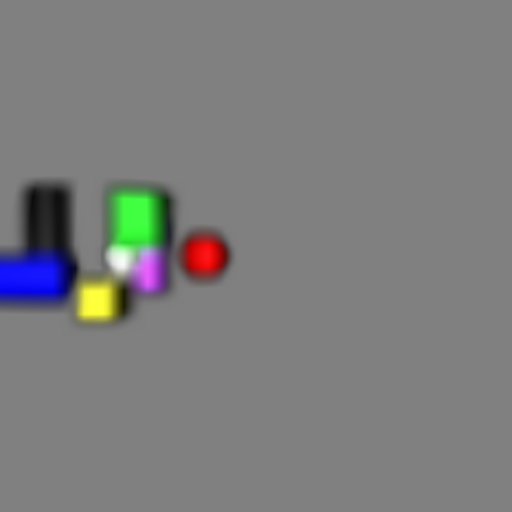} & \includegraphics[width=\linewidth]{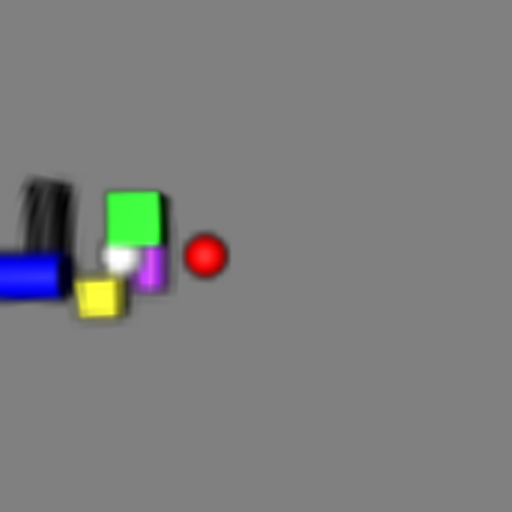} \\
		\hline
    	\resizebox{\linewidth}{!}{
			\pgfmathsetmacro{\posx}{10}
			\pgfmathsetmacro{\posy}{0}
			\pgfmathsetmacro{\posz}{3}
			\pgfmathsetmacro{\dirx}{-3}
			\pgfmathsetmacro{\diry}{0}
			\pgfmathsetmacro{\dirz}{0}
			\tdplotsetmaincoords{60}{110}
	\begin{tikzpicture}[tdplot_main_coords, scale = .3]

		\newcommand{\tdsphere}[5]{%
			\shade[ball color = #5,
					opacity = 1
			] (#3,#1,#2) circle (#4cm);
		}
		
		\newcommand{\tdcylinderVertical}[6]{%
			\begin{scope}[x={(1,0)}]
				\path (1,0,0);
				\pgfgetlastxy{\cylxx}{\cylxy}
				\path (0,1,0);
				\pgfgetlastxy{\cylyx}{\cylyy}
				\path (0,0,1);
				\pgfgetlastxy{\cylzx}{\cylzy}
				\pgfmathsetmacro{\cylt}{(\cylzy * \cylyx - \cylzx * \cylyy)/ (\cylzy * \cylxx - \cylzx * \cylxy)}
				\pgfmathsetmacro{\ang}{atan(\cylt)}
				\pgfmathsetmacro{\ct}{1/sqrt(1 + (\cylt)^2)}
				\pgfmathsetmacro{\st}{\cylt * \ct}
				\shade[ball color = #6, opacity = 1] (#3+#4*\ct,#1+#4*\st,#2+#5/2) -- ++(0,0,-1*#5) arc[start angle=\ang,delta angle=180,radius=#4] -- ++(0,0,#5) arc[start angle=\ang+180,delta angle=-180,radius=#4];
				\begin{scope}[every path/.style={ultra thick}]
					\shade[ball color = #6, opacity = 1] (#3,#1,#2+#5/2) circle[radius=#4];
				\end{scope}
			\end{scope}
		}
		
		\newcommand{\tdcylinderHorizontal}[6]{%
			\begin{scope}[x={(1,0)}]
				\path (1,0,0);
				\pgfgetlastxy{\cylxx}{\cylxy}
				\path (0,1,0);
				\pgfgetlastxy{\cylyx}{\cylyy}
				\path (0,0,1);
				\pgfgetlastxy{\cylzx}{\cylzy}
				\pgfmathsetmacro{\cylt}{(\cylxy * \cylyx - \cylxx * \cylyy)/ (\cylxy * \cylzx - \cylxx * \cylzy)}
				\pgfmathsetmacro{\ang}{atan(\cylt)}
				\pgfmathsetmacro{\ct}{1/sqrt(1 + (\cylt)^2)}
				\pgfmathsetmacro{\st}{\cylt * \ct}
				
				\shade[ball color = #6, opacity = 1] (#3+#5/2,#1-#4*\st,#2-#4*\ct) -- ++(-#5,0,0) {[canvas is yz plane at x=#3-#5/2] arc[start angle=\ang+30,delta angle=180,radius=#4]} -- ++(#5,0,0) {[canvas is yz plane at x=#3+#5/2] arc[start angle=\ang+180+30,delta angle=-180,radius=#4]};
				\begin{scope}[every path/.style={ultra thick}, canvas is yz plane at x=#3+#5/2	]
					\shade[ball color = #6, opacity = 1] (#1,#2,0) circle[radius=#4];
				\end{scope}
			\end{scope}
		}
		
		\newcommand{\tdcube}[7]{%
				\shade[ball color = #7, opacity = 1] (#3-#6/2,#1-#4/2,#2+#5/2) -- ++(#6,0,0) -- ++(0,#4,0) -- ++(-#6,0,0) -- ++ (0,-#4,0);
				\shade[ball color = #7, opacity = 1] (#3+#6/2,#1-#4/2,#2+#5/2) -- ++(0,#4,0) -- ++(0,0,-#5) -- ++(0,-#4,0) -- ++ (0,0,#5);
				\shade[ball color = #7, opacity = 1] (#3+#6/2,#1+#4/2,#2+#5/2) -- ++(-#6,0,0) -- ++(0,0,-#5) -- ++(#6,0,0) -- ++ (0,0,#5);
		}
	 
		\draw[-stealth] (0,0,0) -- (11,0,0) 
				node[below left] {\huge $z$};
		 
		\draw[-stealth] (0,0,0) -- (0,11,0)
				node[below right] {\huge $x$};
		 
		\draw[-stealth] (0,0,0) -- (0,0,11)
				node[above] {\huge $y$};
				
		\draw[dashed] (0,0,0) -- (-11,0,0);
		 
		\draw[dashed] (0,0,0) -- (0,-11,0);
		 
		\draw[dashed] (0,0,0) -- (0,0,-11);

		\coordinate (P) at (\posz,\posx,\posy);
		\coordinate (D) at (\dirz,\dirx,\diry);
		\draw[red,fill=red] (P) circle (.5ex);
		\draw[-{stealth[scale=2]},color=red, very thick] (P) -- ($(P)+(D)$);
		
		\tdsphere{0}{0}{0}{.5}{white}
		\tdsphere{0}{0}{-1.87}{.5}{red}
		\tdcylinderVertical{-2.62}{0}{0}{.5}{2}{violet}
		\tdcylinderHorizontal{1.07}{-.41}{1.81}{.5}{2}{blue}
		\tdcube{-1.47}{-1.05}{0.93}{1}{1}{1}{yellow}
		\tdcylinderVertical{-0.87}{0.76}{2.06}{.5}{2}{black}
		\tdcube{1.2}{0.71}{-0.68}{1}{1}{1}{green}
	 
	\end{tikzpicture}
		} & \includegraphics[height=\linewidth]{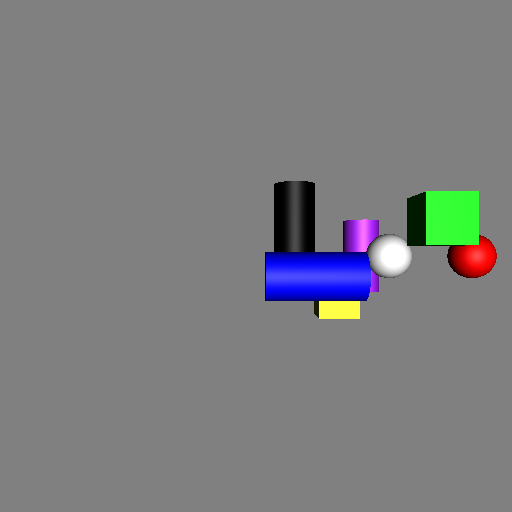} & \includegraphics[width=\linewidth]{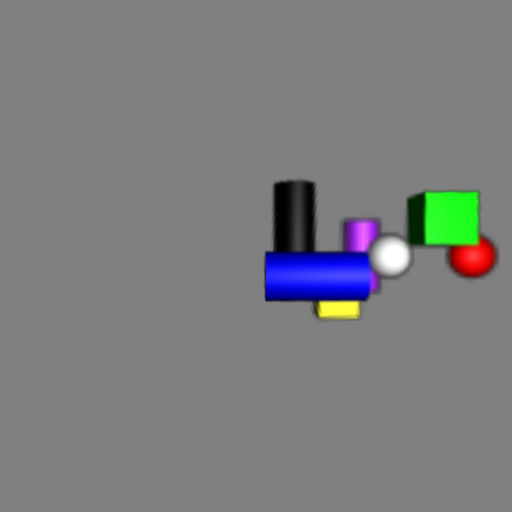} & \includegraphics[width=\linewidth]{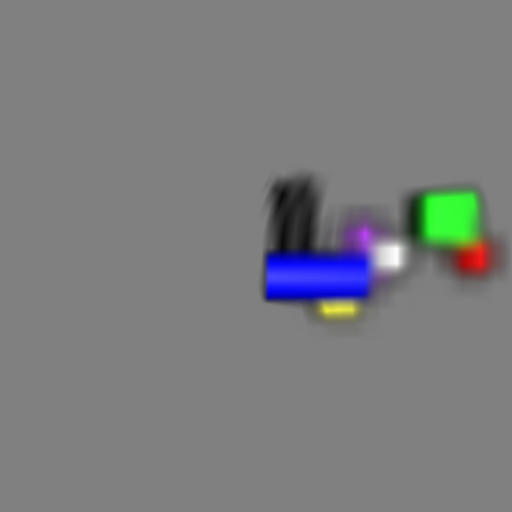}& \includegraphics[width=\linewidth]{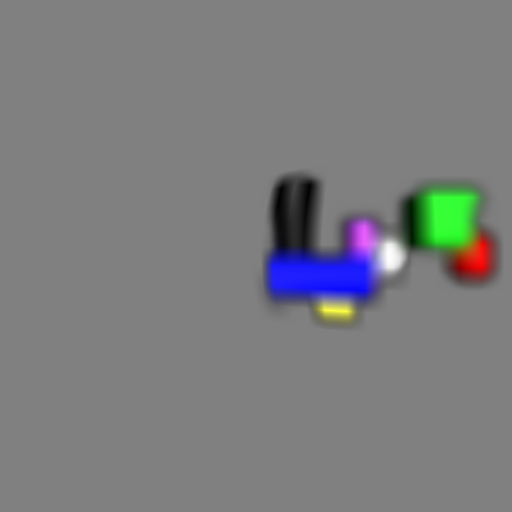} & \includegraphics[width=\linewidth]{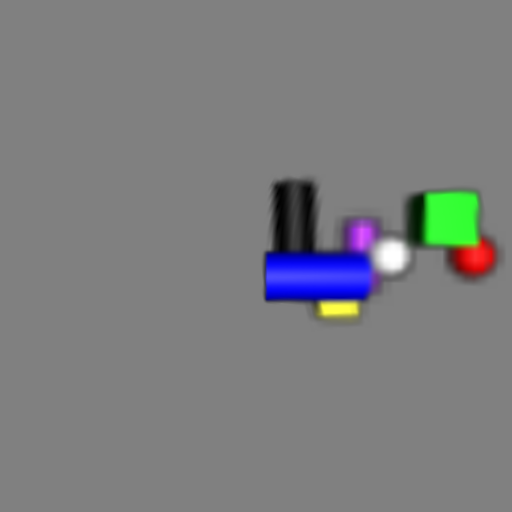}\\
		\hline
    	\resizebox{\linewidth}{!}{
			\pgfmathsetmacro{\posx}{0}
			\pgfmathsetmacro{\posy}{0}
			\pgfmathsetmacro{\posz}{10}
			\pgfmathsetmacro{\dirx}{0}
			\pgfmathsetmacro{\diry}{0}
			\pgfmathsetmacro{\dirz}{-3}
			\tdplotsetmaincoords{60}{110}
	\begin{tikzpicture}[tdplot_main_coords, scale = .3]

		\newcommand{\tdsphere}[5]{%
			\shade[ball color = #5,
					opacity = 1
			] (#3,#1,#2) circle (#4cm);
		}
		
		\newcommand{\tdcylinderVertical}[6]{%
			\begin{scope}[x={(1,0)}]
				\path (1,0,0);
				\pgfgetlastxy{\cylxx}{\cylxy}
				\path (0,1,0);
				\pgfgetlastxy{\cylyx}{\cylyy}
				\path (0,0,1);
				\pgfgetlastxy{\cylzx}{\cylzy}
				\pgfmathsetmacro{\cylt}{(\cylzy * \cylyx - \cylzx * \cylyy)/ (\cylzy * \cylxx - \cylzx * \cylxy)}
				\pgfmathsetmacro{\ang}{atan(\cylt)}
				\pgfmathsetmacro{\ct}{1/sqrt(1 + (\cylt)^2)}
				\pgfmathsetmacro{\st}{\cylt * \ct}
				\shade[ball color = #6, opacity = 1] (#3+#4*\ct,#1+#4*\st,#2+#5/2) -- ++(0,0,-1*#5) arc[start angle=\ang,delta angle=180,radius=#4] -- ++(0,0,#5) arc[start angle=\ang+180,delta angle=-180,radius=#4];
				\begin{scope}[every path/.style={ultra thick}]
					\shade[ball color = #6, opacity = 1] (#3,#1,#2+#5/2) circle[radius=#4];
				\end{scope}
			\end{scope}
		}
		
		\newcommand{\tdcylinderHorizontal}[6]{%
			\begin{scope}[x={(1,0)}]
				\path (1,0,0);
				\pgfgetlastxy{\cylxx}{\cylxy}
				\path (0,1,0);
				\pgfgetlastxy{\cylyx}{\cylyy}
				\path (0,0,1);
				\pgfgetlastxy{\cylzx}{\cylzy}
				\pgfmathsetmacro{\cylt}{(\cylxy * \cylyx - \cylxx * \cylyy)/ (\cylxy * \cylzx - \cylxx * \cylzy)}
				\pgfmathsetmacro{\ang}{atan(\cylt)}
				\pgfmathsetmacro{\ct}{1/sqrt(1 + (\cylt)^2)}
				\pgfmathsetmacro{\st}{\cylt * \ct}
				
				\shade[ball color = #6, opacity = 1] (#3+#5/2,#1-#4*\st,#2-#4*\ct) -- ++(-#5,0,0) {[canvas is yz plane at x=#3-#5/2] arc[start angle=\ang+30,delta angle=180,radius=#4]} -- ++(#5,0,0) {[canvas is yz plane at x=#3+#5/2] arc[start angle=\ang+180+30,delta angle=-180,radius=#4]};
				\begin{scope}[every path/.style={ultra thick}, canvas is yz plane at x=#3+#5/2	]
					\shade[ball color = #6, opacity = 1] (#1,#2,0) circle[radius=#4];
				\end{scope}
			\end{scope}
		}
		
		\newcommand{\tdcube}[7]{%
				\shade[ball color = #7, opacity = 1] (#3-#6/2,#1-#4/2,#2+#5/2) -- ++(#6,0,0) -- ++(0,#4,0) -- ++(-#6,0,0) -- ++ (0,-#4,0);
				\shade[ball color = #7, opacity = 1] (#3+#6/2,#1-#4/2,#2+#5/2) -- ++(0,#4,0) -- ++(0,0,-#5) -- ++(0,-#4,0) -- ++ (0,0,#5);
				\shade[ball color = #7, opacity = 1] (#3+#6/2,#1+#4/2,#2+#5/2) -- ++(-#6,0,0) -- ++(0,0,-#5) -- ++(#6,0,0) -- ++ (0,0,#5);
		}
	 
		\draw[-stealth] (0,0,0) -- (11,0,0) 
				node[below left] {\huge $z$};
		 
		\draw[-stealth] (0,0,0) -- (0,11,0)
				node[below right] {\huge $x$};
		 
		\draw[-stealth] (0,0,0) -- (0,0,11)
				node[above] {\huge $y$};
				
		\draw[dashed] (0,0,0) -- (-11,0,0);
		 
		\draw[dashed] (0,0,0) -- (0,-11,0);
		 
		\draw[dashed] (0,0,0) -- (0,0,-11);

		\coordinate (P) at (\posz,\posx,\posy);
		\coordinate (D) at (\dirz,\dirx,\diry);
		\draw[red,fill=red] (P) circle (.5ex);
		\draw[-{stealth[scale=2]},color=red, very thick] (P) -- ($(P)+(D)$);
		
		\tdsphere{0}{0}{0}{.5}{white}
		\tdsphere{0}{0}{-1.87}{.5}{red}
		\tdcylinderVertical{-2.62}{0}{0}{.5}{2}{violet}
		\tdcylinderHorizontal{1.07}{-.41}{1.81}{.5}{2}{blue}
		\tdcube{-1.47}{-1.05}{0.93}{1}{1}{1}{yellow}
		\tdcylinderVertical{-0.87}{0.76}{2.06}{.5}{2}{black}
		\tdcube{1.2}{0.71}{-0.68}{1}{1}{1}{green}
	 
	\end{tikzpicture}
		} & \includegraphics[height=\linewidth]{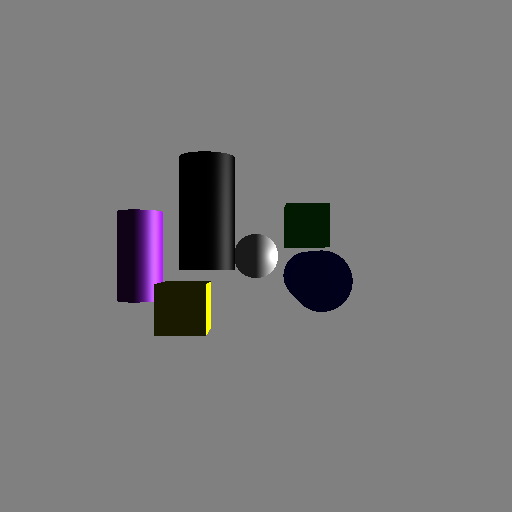} & \includegraphics[width=\linewidth]{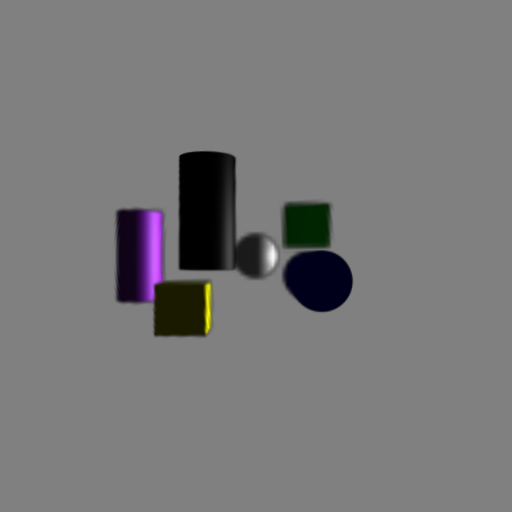} & \includegraphics[width=\linewidth]{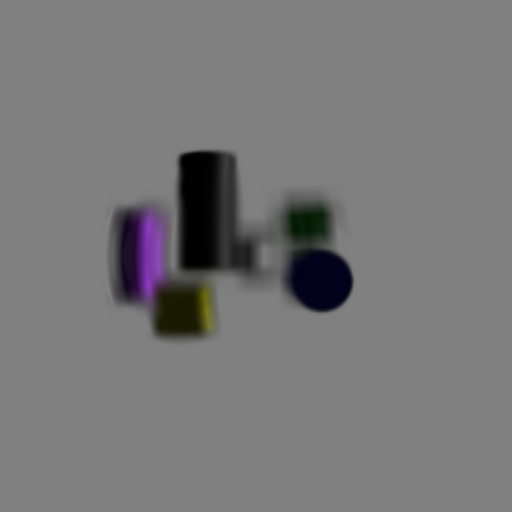}& \includegraphics[width=\linewidth]{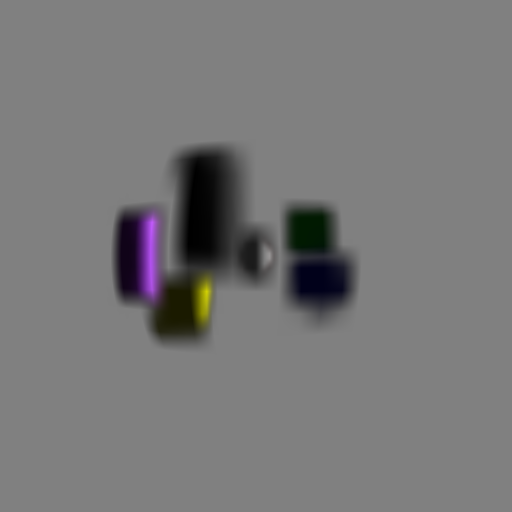} & \includegraphics[width=\linewidth]{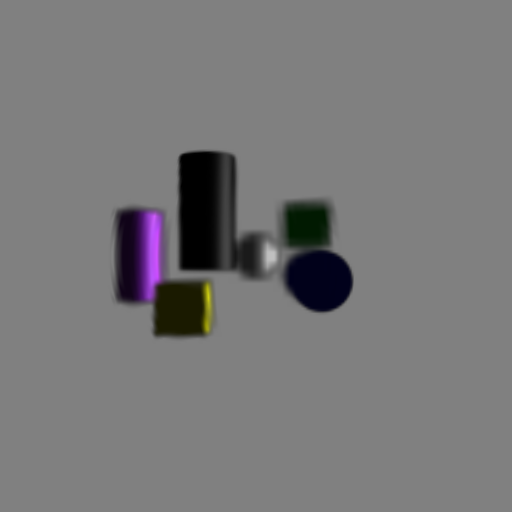}\\
		\hline
    	\resizebox{\linewidth}{!}{
			\pgfmathsetmacro{\posx}{0}
			\pgfmathsetmacro{\posy}{10}
			\pgfmathsetmacro{\posz}{0}
			\pgfmathsetmacro{\dirx}{0}
			\pgfmathsetmacro{\diry}{-3}
			\pgfmathsetmacro{\dirz}{0}
			\tdplotsetmaincoords{60}{110}
	\begin{tikzpicture}[tdplot_main_coords, scale = .3]

		\newcommand{\tdsphere}[5]{%
			\shade[ball color = #5,
					opacity = 1
			] (#3,#1,#2) circle (#4cm);
		}
		
		\newcommand{\tdcylinderVertical}[6]{%
			\begin{scope}[x={(1,0)}]
				\path (1,0,0);
				\pgfgetlastxy{\cylxx}{\cylxy}
				\path (0,1,0);
				\pgfgetlastxy{\cylyx}{\cylyy}
				\path (0,0,1);
				\pgfgetlastxy{\cylzx}{\cylzy}
				\pgfmathsetmacro{\cylt}{(\cylzy * \cylyx - \cylzx * \cylyy)/ (\cylzy * \cylxx - \cylzx * \cylxy)}
				\pgfmathsetmacro{\ang}{atan(\cylt)}
				\pgfmathsetmacro{\ct}{1/sqrt(1 + (\cylt)^2)}
				\pgfmathsetmacro{\st}{\cylt * \ct}
				\shade[ball color = #6, opacity = 1] (#3+#4*\ct,#1+#4*\st,#2+#5/2) -- ++(0,0,-1*#5) arc[start angle=\ang,delta angle=180,radius=#4] -- ++(0,0,#5) arc[start angle=\ang+180,delta angle=-180,radius=#4];
				\begin{scope}[every path/.style={ultra thick}]
					\shade[ball color = #6, opacity = 1] (#3,#1,#2+#5/2) circle[radius=#4];
				\end{scope}
			\end{scope}
		}
		
		\newcommand{\tdcylinderHorizontal}[6]{%
			\begin{scope}[x={(1,0)}]
				\path (1,0,0);
				\pgfgetlastxy{\cylxx}{\cylxy}
				\path (0,1,0);
				\pgfgetlastxy{\cylyx}{\cylyy}
				\path (0,0,1);
				\pgfgetlastxy{\cylzx}{\cylzy}
				\pgfmathsetmacro{\cylt}{(\cylxy * \cylyx - \cylxx * \cylyy)/ (\cylxy * \cylzx - \cylxx * \cylzy)}
				\pgfmathsetmacro{\ang}{atan(\cylt)}
				\pgfmathsetmacro{\ct}{1/sqrt(1 + (\cylt)^2)}
				\pgfmathsetmacro{\st}{\cylt * \ct}
				
				\shade[ball color = #6, opacity = 1] (#3+#5/2,#1-#4*\st,#2-#4*\ct) -- ++(-#5,0,0) {[canvas is yz plane at x=#3-#5/2] arc[start angle=\ang+30,delta angle=180,radius=#4]} -- ++(#5,0,0) {[canvas is yz plane at x=#3+#5/2] arc[start angle=\ang+180+30,delta angle=-180,radius=#4]};
				\begin{scope}[every path/.style={ultra thick}, canvas is yz plane at x=#3+#5/2	]
					\shade[ball color = #6, opacity = 1] (#1,#2,0) circle[radius=#4];
				\end{scope}
			\end{scope}
		}
		
		\newcommand{\tdcube}[7]{%
				\shade[ball color = #7, opacity = 1] (#3-#6/2,#1-#4/2,#2+#5/2) -- ++(#6,0,0) -- ++(0,#4,0) -- ++(-#6,0,0) -- ++ (0,-#4,0);
				\shade[ball color = #7, opacity = 1] (#3+#6/2,#1-#4/2,#2+#5/2) -- ++(0,#4,0) -- ++(0,0,-#5) -- ++(0,-#4,0) -- ++ (0,0,#5);
				\shade[ball color = #7, opacity = 1] (#3+#6/2,#1+#4/2,#2+#5/2) -- ++(-#6,0,0) -- ++(0,0,-#5) -- ++(#6,0,0) -- ++ (0,0,#5);
		}
	 
		\draw[-stealth] (0,0,0) -- (11,0,0) 
				node[below left] {\huge $z$};
		 
		\draw[-stealth] (0,0,0) -- (0,11,0)
				node[below right] {\huge $x$};
		 
		\draw[-stealth] (0,0,0) -- (0,0,11)
				node[above] {\huge $y$};
				
		\draw[dashed] (0,0,0) -- (-11,0,0);
		 
		\draw[dashed] (0,0,0) -- (0,-11,0);
		 
		\draw[dashed] (0,0,0) -- (0,0,-11);

		\coordinate (P) at (\posz,\posx,\posy);
		\coordinate (D) at (\dirz,\dirx,\diry);
		\draw[red,fill=red] (P) circle (.5ex);
		\draw[-{stealth[scale=2]},color=red, very thick] (P) -- ($(P)+(D)$);
		
		\tdsphere{0}{0}{0}{.5}{white}
		\tdsphere{0}{0}{-1.87}{.5}{red}
		\tdcylinderVertical{-2.62}{0}{0}{.5}{2}{violet}
		\tdcylinderHorizontal{1.07}{-.41}{1.81}{.5}{2}{blue}
		\tdcube{-1.47}{-1.05}{0.93}{1}{1}{1}{yellow}
		\tdcylinderVertical{-0.87}{0.76}{2.06}{.5}{2}{black}
		\tdcube{1.2}{0.71}{-0.68}{1}{1}{1}{green}
	 
	\end{tikzpicture}
		} & \includegraphics[height=\linewidth]{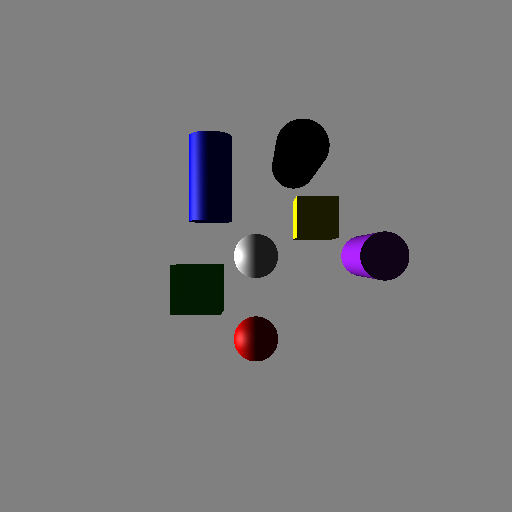} & \includegraphics[width=\linewidth]{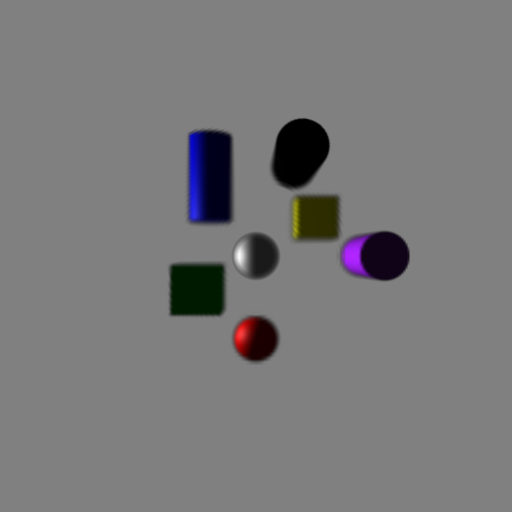} & \includegraphics[width=\linewidth]{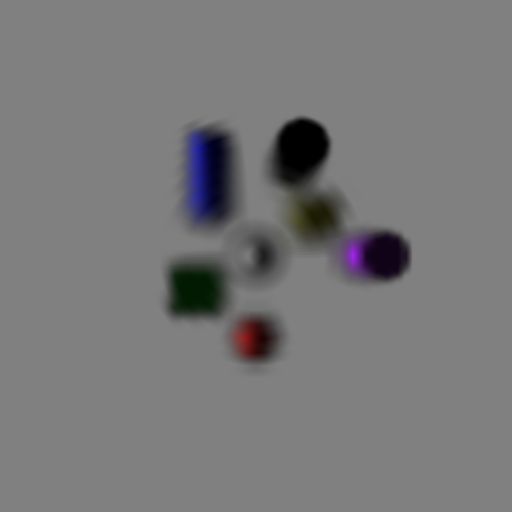}& \includegraphics[width=\linewidth]{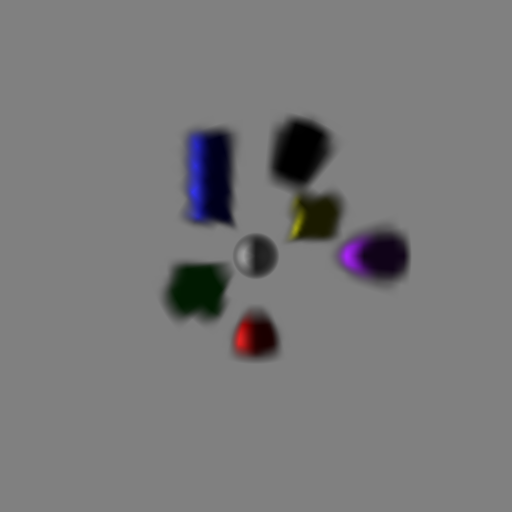} & \includegraphics[width=\linewidth]{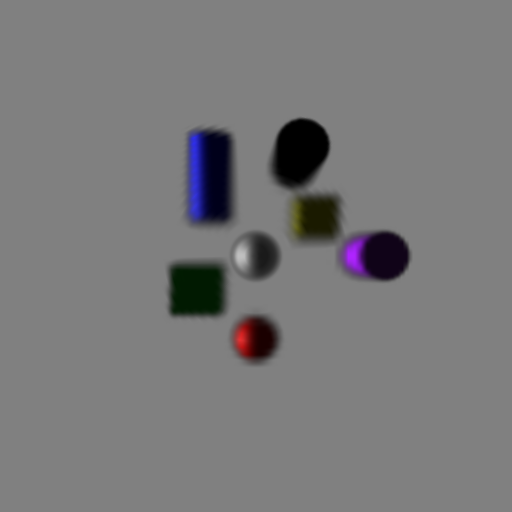}\\
		\hline
  \end{tabular}
  \caption{Comparative between the models' renderization and the effect produced by our texture under several configuration of resolutions over several point of view.}
  \label{tab:comparative_resolution}
\end{figure*}

We also synthesize this same texture under several configuration of angular and spatial resolution and compare them to the direct render of the scenario under several positions and directions (Fig. \ref{tab:comparative_resolution}) to check if the result was coherent with the desired one and to identify which dimensions were more important for the final result.

\section{Results}

Our texture structure was able to simulate the same behaviour observed in the direct render of our objects, including the parallax effect as desired. 
\par Naturally when the resolution of the texture increases, the quality of the output image also increase. A well balance between the spatial and angular resolution is essential for the overall quality of an image and to avoid aliasing effects.
\par By comparing the output images showed in fig. \ref{tab:comparative_resolution} we observed that the spatial dimensions had larger contribution to the overall format representation of the visible objects. This effect is more expressive in the areas found with more irregular grids (as happens over the poles of a spherical parametrization) and for the format definition of the objects close to the part of the viewable model surface, while the angular resolution had larger contribution to the definition of objects far from this surface's part and to the sharpness of corners and borders.
\par It's worth mentioning the aliasing effect caused by the low resolution over the spatial dimension. This aliasing effect is hard to be showed in images and is more noticeable when the observer is moving tangentially to the model surface. The objects projection over our surface appears to be moving in waves, in other words, as the observer moves, some points appears to be statical and others in movements, the behaviour of these points switchs as the observer keeps moving.
\par A benchmark test on Unity using a virtual environment with only a sphere where our texture of dimension 512x256x32x32 were applied was able to run at 180 to 200 fps, showing that our structure is applicable to real time system. This benchmark test was done using a computer with a processor i7-8750H and a GPU RTX2070.

\section{Conclusion}

This work proposes a new texture structure that successfully simulate parallax effect in objects in a virtual world through the simulation of light rays emitted by them. Beyond the simulation of parallax, it is capable of reproducing any effect included on it during its synthesis including shadows and reflection. The addition of layers of effects during the render, such as diffuse and specular reflections, were not studied, but theoretically it is possible if the model follows its particular conditions. This structure needs low process power, what made possible to implement it on real time system, but requires a large storage capacity. Ways of compacting it were not studied.
\par This work explores only one way of rendering our textures and other may be explored. One possible suggestion would be to change the scale of the angular dimensions to emphasize the direct angle view over the tangential ones.
\par Also the defocusing effect observed on our texture is very similar to the one observed in photographic machines. As an analogy, we can compare our spheric surface to the focal plane and the angular resolution to the aperture size. As a next work we intend to explore this relationship and synthesize a texture after a series of photos.
\par We also intend to explore the structure developed here to improve our spheric display to a true multi-viewer experience with full parallax perception and eliminate the restriction of simultaneous users.

\section{Acknowledgment}

We would like to express our sincere gratitude to the Laboratory of Integrable Systems (LSI) and the Interdisciplinary Center in Interactive Technologies (CITI) for their invaluable support and collaboration throughout this research. Additionally, we extend our thanks to the colleagues Luis G.F. da Costa, Mario R. Nagamura, Celso S. Kurashima, Matteus Car, Arthur A. Miyazaki and Lucas R. Mata for their valuable contributions and enriching daily conversations, which have greatly influenced this paper.

\bibliographystyle{plain}
\bibliography{ArtigoTextura2+2}

\end{document}